      \documentstyle[12pt]{article}
    \title{{\bf  A theory of tensor products for module categories for
a vertex operator algebra, I}}
    \author{Yi-Zhi Huang and James Lepowsky}
    \date{}
    
\begin{document}

    \hyphenation{Phil-a-del-phia equip-ped}
    \bibliographystyle{alpha}
    \maketitle

    \input amssym.def
    \input amssym
    \newtheorem{rema}{Remark}[section]
    \newtheorem{propo}[rema]{Proposition}
    \newtheorem{theo}[rema]{Theorem}
   \newtheorem{defi}[rema]{Definition}
    \newtheorem{lemma}[rema]{Lemma}
    \newtheorem{corol}[rema]{Corollary}
     \newtheorem{exam}[rema]{Example}
	\newcommand{\ba}{\begin{array}}
	\newcommand{\ea}{\end{array}}
        \newcommand{\be}{\begin{equation}}
        \newcommand{\ee}{\end{equation}}
	\newcommand{\bea}{\begin{eqnarray}}
	\newcommand{\eea}{\end{eqnarray}}
	\newcommand{\nno}{\nonumber}
	\newcommand{\lbar}{\bigg\vert}
	\newcommand{\p}{\partial}
	\newcommand{\dps}{\displaystyle}
	\newcommand{\bra}{\langle}
	\newcommand{\ket}{\rangle}
 \newcommand{\res}{\mbox{ \rm Res}}
 \newcommand{\pf}{{\it Proof}\hspace{2ex}}
 \newcommand{\epf}{\hspace{1em}$\Box$}
 \newcommand{\epfv}{\hspace{1em}$\Box$\vspace{1em}}

\makeatletter
\newlength{\@pxlwd} \newlength{\@rulewd} \newlength{\@pxlht}
\catcode`.=\active \catcode`B=\active \catcode`:=\active \catcode`|=\active
\def\sprite#1(#2,#3)[#4,#5]{
   \edef\@sprbox{\expandafter\@cdr\string#1\@nil @box}
   \expandafter\newsavebox\csname\@sprbox\endcsname
   \edef#1{\expandafter\usebox\csname\@sprbox\endcsname}
   \expandafter\setbox\csname\@sprbox\endcsname =\hbox\bgroup
   \vbox\bgroup
      \catcode`.=\active\catcode`B=\active\catcode`:=\active\catcode`|=\active
      \@pxlwd=#4 \divide\@pxlwd by #3 \@rulewd=\@pxlwd
      \@pxlht=#5 \divide\@pxlht by #2
      \def .{\hskip \@pxlwd \ignorespaces}
      \def B{\@ifnextchar B{\advance\@rulewd by \@pxlwd}{\vrule
         height \@pxlht width \@rulewd depth 0 pt \@rulewd=\@pxlwd}}
      \def :{\hbox\bgroup\vrule height \@pxlht width 0pt depth
0pt\ignorespaces}
      \def |{\vrule height \@pxlht width 0pt depth 0pt\egroup
         \prevdepth= -1000 pt}
   }
\def\endsprite{\egroup\egroup}
\catcode`.=12 \catcode`B=11 \catcode`:=12 \catcode`|=12\relax
\makeatother

\def\hboxtr{\FormOfHboxtr} 
\sprite{\FormOfHboxtr}(25,25)[0.5 em, 1.2 ex] 

:BBBBBBBBBBBBBBBBBBBBBBBBB |
:BB......................B |
:B.B.....................B |
:B..B....................B |
:B...B...................B |
:B....B..................B |
:B.....B.................B |
:B......B................B |
:B.......B...............B |
:B........B..............B |
:B.........B.............B |
:B..........B............B |
:B...........B...........B |
:B............B..........B |
:B.............B.........B |
:B..............B........B |
:B...............B.......B |
:B................B......B |
:B.................B.....B |
:B..................B....B |
:B...................B...B |
:B....................B..B |
:B.....................B.B |
:B......................BB |
:BBBBBBBBBBBBBBBBBBBBBBBBB |

\endsprite

\tableofcontents

\begin{abstract}

This is the first part in a series of papers developing a tensor
product theory for modules for a vertex operator algebra.  The goal of
this theory is to construct a ``vertex tensor category'' structure on
the category of modules for a suitable vertex operator algebra.  The
notion of vertex tensor category is essentially a ``complex analogue''
of the notion of symmetric tensor category, and in fact a vertex
tensor category produces a braided tensor category in a natural way.
The theory applies in particular to many familiar ``rational'' vertex
operator algebras, including those associated with WZNW models,
minimal models and the moonshine module.  In this paper (Part I), we
introduce the notions of $P(z)$- and $Q(z)$-tensor product, where
$P(z)$ and $Q(z)$ are two special elements of the moduli space of
spheres with punctures and local coordinates, and we present the
fundamental properties and constructions of $Q(z)$-tensor products.

\end{abstract}

\renewcommand{\theequation}{\thesection.\arabic{equation}}
\renewcommand{\therema}{\thesection.\arabic{rema}}
\setcounter{equation}{0}
\setcounter{rema}{0}

\section{Introduction}

In this paper, we begin the detailed development of the theory of
tensor products of modules for a vertex operator algebra initiated in
\cite{HL1} and described in \cite{HL5}. Because this work is based on
a new point of view and new techniques, we start with some fairly
extensive motivation for and explanation of this point of view before
presenting the main content of this paper.

It has become more and more evident that the theory of vertex operator
algebras and their representations provides the natural foundation and
context for the deeper study of a wide range of structures and
concepts, including the Fischer-Griess Monster sporadic finite simple
group and its relations with ``monstrous moonshine"; conformal field
theory as it occurs in both mathematics and physics; representation
theory of several classes of infinite-dimensional Lie algebras,
including affine Kac-Moody algebras and the Virasoro algebra; and
topological invariants such as the Jones \cite{J}, HOMFLY and Jones-Witten
\cite{W2} invariants.  The theory, which has been extensively developed in
many directions and under different guises, has exhibited a striking
philosophical paradox, which may be described as follows (cf. \cite{L}):

One of the main motivations for the development of the theory in
\cite{B1} and \cite{FLM2} was the need for a natural
infinite-dimensional structure on which the Monster would act in a
natural way as the automorphism group, and whose properties would
provide a proof of the Conway-Norton conjectures \cite{CN} relating
the Monster to modular functions.  In fact, a certain part of these
conjectures was proved in \cite{FLM1} and \cite{FLM2} by means of the
construction of the ``moonshine module" and of the action of the
Monster on it, and the rest of the main conjecture was established in
\cite{B2}, where the McKay-Thompson series for the moonshine module
not determined in \cite{FLM1}, \cite{FLM2} were all
determined and found to agree with those proposed in \cite{CN}.

The Monster is arguably the most ``exceptional" finite symmetry group
that nature allows, and while there is no claim that it is easy to
define, the moonshine module is arguably the most natural structure on
which the Monster acts as the automorphism group.  On the other hand,
in a certain precise sense, this structure which has the richest
possible symmetry group exhibits the most trivial possible monodromy:
In conformal field theory or vertex operator algebra theory, the
monodromy of the correlation functions formed {}from intertwining
operators among modules for a suitable vertex operator algebra
provides representations of braid groups and other mapping class
groups.  These in turn lead to the construction of certain braided
monoidal (tensor) categories (see \cite{JS}) and to knot and
3-manifold invariants, for such conformal field theories as the
Wess-Zumino-Novikov-Witten model \cite{W1}.  One can also obtain these
braided monoidal categories {}from representations of certain quantum
groups.  For important aspects of this extensive program, see in
particular \cite{BPZ}, \cite{Dr1}, \cite{Dr2},
\cite{KZ}, \cite{K1}, \cite{K2},
\cite{MS}, \cite{RT}, \cite{SV},
\cite{TK}, \cite{Va}.  But as conjectured in \cite{FLM2} and proved by Dong
\cite{Do},
the moonshine module vertex operator algebra admits only itself as an
irreducible module, and this implies the monodromy-triviality; that
is, in terms of monodromy, the moonshine module is no different {}from
the {\it trivial} conformal field theory.  The subtlety of the Monster
and the moonshine module is ``orthogonal" to the subtlety of braiding
and fusing in conformal field theory (cf.
\cite{MS}).  Thus we have the fundamental question of ``lifting" both
the theory of the symmetry-rich but monodromy-free model, the
moonshine module, and also the theory of the monodromy-rich
models, to a common theory incorporating the deepest features of
both.

On the other hand, in recent years, there has been much effort by
mathematicians to develop conformal field theory as a serious
mathematical theory.  Certain works by physicists already suggested
that if the solid foundation of conformal field theory could be laid
down, there would be many interesting and powerful mathematical
consequences and physical applications.  Segal's definition of
conformal field theory \cite{Se} was essentially the first work by a
mathematician in this direction.  Though his definition is beautiful
and rigorous, the problem of constructing even a single complete
nontrivial example satisfying the definition has remained open. Most
of the mathematical works on conformal field theories center on
``modular functors'' or ``conformal blocks'' (see for example
\cite{Se},
\cite{TUY}), including monodromy and representations of braid groups
and other mapping class groups, rather than on conformal fields
themselves.  The difficulties involve not just some new formalism
needed to rigorize physicists' work, but more importantly, certain
powerful machinery enabling one to construct a conformal field theory
{}from some relatively elementary mathematical data.  The theory of
vertex operator algebras provides exactly the formalism we need, and
the appropriate powerful machinery is being developed within its
framework.

In the present series of papers, an essential part of the new
formalism and machinery, a tensor product theory for modules for a
vertex operator algebra, will be developed, with the problems and
requirements above as its most basic goals and features.  The main
idea is to exploit fully the {\it conformal} structure of conformal
field theory rather than just the {\it topological} structure upon
which the familiar considerations of monodromy and braided monoidal
categories are based.  The precise formulation of this idea is
accomplished with the help of the notion of operad, which was
originally introduced in a topological context (\cite{St1},
\cite{St2}, \cite{M})
in connection with the homotopy-theoretic characterization of loop
spaces.  Operads are systems which abstract the notion and properties
of such operations as substitution operations in algebra or certain
sewing operations in geometry.  They can be found ``everywhere":
Classical algebraic structures like groups, algebras and Lie algebras
are in fact always implicitly based on operads defined using
one-dimensional geometric objects such as punctured circles and binary
trees.  In \cite{HL2} and \cite{HL3}, the geometric interpretation
given by the first-named author (\cite{H1}, \cite{H2}, \cite{H5}) of
the notion of vertex operator algebra is used to define a certain
operadic structure, but this time two-dimensional (or rather,
one-complex-dimensional), based on spheres with tubes (punctures and
local coordinates), in such a way that vertex operator algebras can be
viewed as based on these structures by analogy with the way that say
associative algebras can be viewed as based on the circle with
punctures and local coordinates. This amounts essentially to a precise
formulation of the well-known philosophy of the physical and geometric
foundation of conformal field theory at genus zero. For the exact
definition and the detailed study of this particular operadic
structure, the ``vertex (partial) operad,'' the reader is referred to
\cite{HL2}, \cite{HL3} and \cite{H5}.

This operadic formulation of the notion of vertex operator algebra
enables us to raise questions at a fundamental level: For instance,
the very notion of tensor category can be viewed as based on the
circle operad.  Is there a reasonable ``vertex" analogue of the notion
of tensor category, based on the ``vertex operad" of \cite{HL2},
\cite{HL3}?  An answer to this question is given in \cite{HL5}, an
overview of the theory being developed in this series of papers, in
the form of a new notion of ``vertex tensor category"---an abelian
category equipped with suitably symmetric ``tensor product''
operations and associativity isomorphisms and constraints, etc., which
are essentially parametrized and controlled by spheres with punctures
and local coordinates rather than by the traditional natural
isomorphisms and commutative diagrams of coherence theory for monoidal
categories.  This is not merely a curious abstraction; the goal of the
present series of papers is to show that the category of modules for a
suitable vertex operator algebra is a vertex tensor category in a
natural way.  Vertex tensor categories contain canonical conformal
information, not just the much more resticted topological (monodromy)
information on which the braided tensor categories are based. In
general, vertex tensor categories naturally give rise to braided
tensor categories, when one forgets the conformal structure and keeps
only the topological information.

In the representation theory of Lie algebras, we have the classical
notion of tensor product of two modules, providing the conceptual
foundation of the Clebsch-Gordan coefficients. The tensor product
operation is an operation on the category of modules for a Lie
algebra, giving a classical example of a tensor category satisfying an
additional symmetry axiom.  For quantum groups (Hopf algebras), the
module categories are also tensor categories but in general do not
satisfy the symmetry axiom, corresponding to the fact that the Hopf
algebra need not be cocommutative. Instead, such tensor categories
satisfy weaker conditions --- braiding conditions. {From} the
resulting braid group representations, one can construct knot and link
invariants. See in particular \cite{J}, \cite{K1}, \cite{K2},
\cite{Dr1}, \cite{Dr2}, \cite{RT}.

Vertex operator algebras (\cite{B1}, \cite{FLM2}, \cite{FHL}) are
``complex analogues'' of both Lie algebras and commutative associative
algebras. They are essentially equivalent to ``chiral algebras'' in
conformal field theory (see in particular \cite{BPZ} and \cite{MS}).
For vertex operator algebras, we also have the notions of modules,
intertwining operators among triples of modules and fusion rules
(dimensions of spaces of intertwining operators) analogous to those
for Lie algebras. We need to use the versions of these notions given in
\cite{FLM2} and \cite{FHL}, and recalled below. In particular,
the appropriate notion of intertwining operator is the one defined in
\cite{FHL}, based on the Jacobi identity axiom for vertex operator
algebras (\cite{FLM2}, \cite{FHL}). In the study of rational conformal
field theories (\cite{BPZ}, \cite{FS}), intertwining operators (or
chiral vertex operators) are fundamental tools.  Many important
concepts and results, for example, representations of braid groups,
the relationship between modular transformations and fusion rules, and
duality relations, are obtained through the study of intertwining
operators; see for example
\cite{KZ},
\cite{TK}, \cite{V} and \cite{MS}. The fusion
rules for a vertex operator algebra being the analogues of the
Clebsch-Gordan coefficients for a Lie algebra, we have the natural
question whether there exists a conceptual notion of tensor product
for modules for a vertex operator algebra which would naturally
provide a conceptual foundation for fusion rules and intertwining
operators.

We noticed a few years ago that the Jacobi identity axiom (see
\cite{FLM2}, \cite{FHL}) for vertex
operator algebras suggests a kind of ``complex analogue'' of the
coalgebra diagonal map for primitive elements of a Hopf algebra, but
it turned out that a considerable amount of work was needed to make
this idea precise and useful. In this paper we begin this program in
detail.  Given two modules $W_{1}$ and $W_{2}$ for a vertex operator
algebra $V$, when one tries to define a tensor product module, the
first serious problem is that the tensor product vector space
$W_{1}\otimes W_{2}$ is not a $V$-module in any natural way (although
it is a module for the vertex operator algebra $V\otimes V$), and so
the underlying vector space of a tensor product module would not be
expected to be the tensor product vector space. Another serious
problem is that a vertex operator algebra is not a Hopf algebra in any
natural sense. We need a new way to define and construct a tensor
product module --- both the underlying vector space and the action of
the vertex operator algebra.  As we shall see, the analogy between
vertex operator algebras and Lie algebras, centered on the Jacobi
identity axiom, provides an analogue of a Hopf algebra diagonal map
for a construction of a tensor product module, under appropriate
hypotheses. In addition, the analogy between vertex operator algebras
and commutative associative algebras, via the geometric and operadic
formulation of the notion of vertex operator algebra (\cite{H1},
\cite{H2}, \cite{HL2}, \cite{HL3}, \cite{H5}),
provides the geometric foundation for the construction.

One important class of examples of vertex operator algebras is
constructed {from} certain modules for affine Lie algebras (see for
example \cite{FZ}, \cite{DL}). There are interesting relations between
representations of affine Lie algebras and of quantum groups discussed
in several of the works mentioned above, for example, and to
understand these relations on a deeper level, one natural strategy is
to compare categories of modules for affine Lie algebras with a fixed
nonzero central charge and categories of modules for associated
quantum groups. While the category of modules for a quantum group is a
tensor category, a category of modules of a fixed nonzero level for an
affine Lie algebras does not close under the classical tensor product
of Lie algebra modules. Thus an appropriate tensor product module, if
it exists, could not be the ordinary one.

Recently, Kazhdan and Lusztig (\cite{KL1}--\cite{KL5}) have found and
constructed such a tensor product operation for certain module
categories of a fixed but (mostly) negative level for an affine Lie
algebra and have shown that these module categories can in fact be
made into braided tensor categories.  Moreover, they have shown that
these tensor categories are equivalent to suitable categories of
modules for corresponding quantum groups.  On the other hand, {}from
the viewpoint of conformal field theory, the more relevant cases
involve positive integral levels, including the case of the category
whose objects are finite direct sums of modules isomorphic to standard
(integrable highest weight) modules of a fixed positive integral level
for an affine Lie algebra. Such ``fusion categories'' have been
discussed on a physical level of rigor in many works, including
\cite{MS}.  Under the assumption, which is a consequence of the theory
developed in the present series of papers, that these categories have
natural braided tensor category structure, Finkelberg \cite{F} has
shown that these braided tensor categories are related to
corresponding braided tensor categories constructed by Kazhdan and
Lusztig, and that they are equivalent to certain categories of
representations of quantum groups. (See also \cite{Va}). The
construction of Kazhdan-Lusztig was in fact motivated by conformal
field theory, and we expected that their tensor product operation
should come {from} more general and natural structures in conformal
field theory.

In \cite{HL1}, partly motivated by the analogy between vertex operator
algebras and Lie algebras and partly motivated by the announcement
\cite{KL1}, a project toward a theory of tensor products for modules
for a vertex operator algebra was initiated. In the present series of
papers, we shall present this theory of tensor products for suitable
module categories for a vertex operator algebra. Our methods are
independent of the methods of \cite{KL1}--\cite{KL5}, even in the case
in which our vertex operator algebra is associated to an affine Lie
algebra.  In place of the braided monoidal categories that arise
{}from the Kazhdan-Lusztig construction (\cite{KL1}--\cite{KL5}), the
result is instead the ``vertex tensor categories'' mentioned above.
What we have is a conceptual ``complexification" of the notion of
symmetric tensor category.  Moreover, as we have mentioned, a
systematic specialization process yields an ordinary braided monoidal
category {}from the vertex tensor category, for suitable vertex
operator algebras.  This category is the usual braided monoidal
category giving the connection with the representation theory of
quantum groups (see
\cite{KL1}--\cite{KL5},
\cite{F}) and knot and link invariants.  That is, the familiar and
fundamental topological information generated by conformal field
theory at genus zero now becomes a specialization of a theory
systematically ``complexified" starting {}from its foundations.  In
the case that the vertex operator algebra is constructed {}from an
affine Lie algebra, the braided tensor category obtained {}from the
vertex tensor category of modules for this vertex operator algebra
gives us in particular a braided tensor category structure on the
relevant category of modules for the corresponding affine Lie algebra;
thus the present theory of tensor products for module categories for a
vertex operator algebra yields expected but nontrivial
conformal-field-theoretical properties of affine Lie algebras {}from a
general viewpoint.

We emphasize that the theory being developed in the present series of
papers is based on the concepts of vertex operator algebra theory
rather than on the methods of \cite{KL1}--\cite{KL5}, which use special
properties of affine Lie algebras, and that an important feature of
this distinction in viewpoints is that the notion of intertwining
operator that is the starting point of this work is the notion based
on the Jacobi identity rather than the notion of intertwining operator
used by many researchers in conformal field theory, based on certain
Lie algebra coinvariants.  The two notions are indeed equivalent for
the WZNW model and related models, but at every stage in the analysis
it is much more natural for us to work with the Jacobi identity for
vertex operator algebras.  Using this Jacobi identity, a canonical
notion of tensor product of modules for a suitable vertex operator
algebra is introduced, defined in terms of an appropriate universal
property and depending on a given element of the moduli space of
spheres with three tubes.  One aspect of the subtle nature of this
construction is that, as was also the case in
\cite{KL1}--\cite{KL5}, the underlying vector
space of the tensor product module is not at all the tensor product
vector space of the given modules.

However, the theory does in fact provide analogues of the concrete
elements of classical tensor product modules (the usual linear
combinations of tensors of elements of the given modules). Let $W_{1}$
and $W_{2}$ be modules for a suitable vertex operator algebra.  For
$w_{(1)}\in W_{1}$ and $w_{(2)}\in W_{2}$, we have the element
$w_{(1)}\boxtimes w_{(2)}$ (more precisely,
$w_{(1)}\boxtimes_{P(z)}w_{(2)}$ or $w_{(1)}\boxtimes_{Q(z)} w_{(2)}$
for $z\in {\Bbb C}^{\times}$ or more generally,
$w_{(1)}\boxtimes_{\tilde{Q}}w_{(2)}$ for any $\tilde{Q}$ in the
determinant line bundle over the moduli space of spheres with
punctures and local coordinates). But these analogues of classical
tensor product elements are elements of the algebraic completion of
the tensor product module $W_{1}\boxtimes W_{2}$ (more precisely,
$W_{1}\boxtimes_{P(z)}W_{2}$ or $W_{1}\boxtimes W_{2}$ or more
generally, $W_{1}\boxtimes_{\tilde{Q}}W_{2}$), not of the tensor
product module itself.  This necessary feature makes the theory much
more difficult than the classical theory.  But this difficulty is
overcome by a characterization of a certain abstractly-defined
subspace of the dual space of the tensor product vector space of two
modules by means of a certain list of conditions, the most important
of which is what we call the ``compatibility condition," which is
motivated by the Jacobi identity and which allows the abstract
machinery to work.  In fact, one of the main theorems---Theorem
6.3---of the present paper says that this space of vectors is in fact
a module in a certain generalized sense; the proof of this result
requires intricate formal calculations based on the Jacobi identity.
This general theorem enables us to establish the conceptual
vertex-tensor-categorical properties of the tensor product operation,
by analogy with the way in which one's ability to write down concrete
tensor product vectors in a classical tensor product module enables
one to establish the properties of classical tensor categories (such
as the associativity or commutativity properties).  Even though the
tensor product of two modules can still be constructed without Theorem
6.3, this result is crucial for proving all the important theorems,
including the associativity (see Part IV
\cite{H6}) and the vertex tensor category structure. The methods in
the present theory are necessarily based heavily on both the machinery
of the purely algebraic formal calculus (see \cite{FLM2} and
\cite{FHL}) and the machinery of the geometric interpretation of the
notion of vertex operator algebra (see
\cite{H1}, \cite{H2}, \cite{H5}).

It is important to note that this theory is both valid and nontrivial
for such vertex operator algebras as the moonshine module.  Even
though the moonshine module exhibits no monodromy, it is expected to
possess rich vertex-tensor-categorical structure coming {}from the
conformal geometry.  In particular, the present theory is expected to
provide a resolution of the philosophical paradox described above.

This theory also reduces the construction of the genus-zero chiral
parts of ``rational conformal field theories'' to the construction of
the corresponding vertex operator algebras, the proof of the
rationality of these vertex operator algebras and the proof of certain
properties of products of intertwining operators for the vertex
operator algebras; these are typically much easier problems.  For some
familiar conformal field theories, for example, the WZNW models and
the minimal models, these problems concerning the corresponding vertex
operator algebras and intertwining operators can be solved easily
using existing results.  For some other interesting conformal field
theories, for example, those constructed {}from certain ${\cal
W}$-algebras and orbifold theories, these problems can also be solved.

In particular, in an application related to both monstrous moonshine
and the construction of conformal field theories, in the case of the
${\Bbb Z}_{2}$-orbifold conformal field theory which produces the
moonshine module for the Monster (as constructed in \cite{FLM2}) the
whole genus-zero chiral part of the theory has recently been shown by
the first-named author \cite{H7}, heavily using the present new tensor
product theory, to be an abelian intertwining algebra, in the sense of
\cite{DL}. (Abelian intertwining algebras are certain generalizations
of vertex operator algebras for which one-dimensional braid group
representations are incorporated naturally into the structure of the
algebra.) Without the tensor product theory, this would be an
exceedingly difficult problem, since, for example, even the proof by
Frenkel, Meurman and the second-named author
\cite{FLM2}
that the moonshine module is a vertex operator algebra is already so
involved.

It should be emphasized that our theory applies (at least) to an
arbitrary rational vertex operator algebra satisfying certain
additional conditions including some convergence conditions.  It
applies in particular to the WZNW models, minimal models and the
moonshine module vertex operator algebra (whose rationality has been
proved by Dong \cite{Do}). Many of the notions, constructions and
techniques also apply to more general vertex operator algebras.

In the remainder of this Introduction, we focus more technically on
the material in this paper. We use the analogy between vertex operator
algebras and Lie algebras as a guide.  In the theory of Lie algebras
we have the following standard notion of intertwining map (of type
${W_{3}}\choose{W_{1}W_{2}}$) among modules $W_1,W_2,W_3$ for a Lie
algebra $V$, with corresponding actions $\pi_1,\pi_2,\pi_3$ of $V$: a
linear map $I$ {from} the tensor product vector space $W_{1}\otimes
W_{2}$ to $W_{3}$, satisfying the identity
\be
\pi_{3}(v)I(w_{(1)}\otimes w_{(2)}) =I(\pi_{1}(v)w_{(1)}\otimes
w_{(2)})+I(w_{(1)}\otimes \pi_{2}(v)w_{(2)})
\ee
for $v\in V$, $w_{(1)}\in W_1$, $w_{(2)}\in W_{2}$.  This ``Jacobi
identity for intertwining maps'' agrees with the Jacobi identity for
$V$ when all three modules are the adjoint module.  Let us call a {\it
product} of $W_{1}$ and $W_{2}$ a third module $W_{3}$ equipped with
an intertwining map $I$ of type ${W_{3}}\choose{W_{1}W_{2}}$; we
denote this by $(W_{3}, I)$.  Then a tensor product of $W_{1}$ and
$W_{2}$ is a product $(W_{1}\otimes W_{2}, \otimes)$ such that given
any product $(W_{3}, I)$, there exists a unique module map $\eta$
{from} $W_{1}\otimes W_{2}$ to $W_{3}$ such that
\begin{equation}
I=\eta\circ \otimes.
\end{equation}
Thus any tensor product of two given modules has the following
property: The intertwining maps {from} the tensor product vector space
of the two modules to a third module correspond naturally to the
module maps {from} the tensor product module to the third module.
Moreover, this universal property characterizes the tensor product
module up to unique isomorphism.

In this paper (Part I), we analogously define notions of $P(z)$-tensor
product and $Q(z)$-tensor product of two modules for a vertex operator
algebra, where $z$ is a nonzero complex number and $P(z)$ and $Q(z)$
are two particular elements, depending on $z$, of a certain moduli
space of spheres with punctures and local coordinates (see
\cite{H1}, \cite{H2}, \cite{H3}, \cite{H4} or \cite{H5}).
We give two constructions of a $Q(z)$-tensor product when the vertex
operator algebra that we consider is such that its module category (or
some fixed subcategory) is closed under a certain operation. This
occurs in particular if the module category of a vertex operator
algebra satisfies certain finiteness and semisimplicity conditions,
and so the $Q(z)$-tensor product of two modules exists in this case.
Such vertex operator algebras are said to be ``rational.''  (For such
algebras we also give a ``tautological'' construction of a tensor
product module; this in fact provides an existence proof.) The
construction of a $P(z)$-tensor product will be given in Part III
\cite{HL6} using the results in the constructions of the $Q(z)$-tensor
product constructed in this paper (Part I). The first of our two
constructions of a $Q(z)$-tensor product is straightforward and
conceptually simple, but it is difficult to use. Our second, much more
useful, construction, presents the $Q(z)$-tensor product module of two
modules $W_{1}$ and $W_{2}$ (when it exists) in terms of the subspace
of the dual $(W_{1}\otimes W_{2})^{*}$ of the vector space tensor
product consisting of the elements satisfying a certain set of
conditions, the most important of which is the ``compatibility
condition.''

The dependence of the tensor product operation on the nonzero complex
number $z$ is a fundamental feature of our theory. The number $z$
actually represents an element in the moduli space of spheres with
punctures and local coordinates mentioned above. In one of the papers
in this series, we shall see that for every element of this moduli
space (more precisely, for any element of the determinant line bundle
over this moduli space (see \cite{H5})), we have a tensor product
operation. The associativity, commutativity and coherence properties
of this tensor product depend on (the determinant line bundle over)
this moduli space and the sewing operation in a natural way.  Such
properties are in fact the data and axioms in the definition of vertex
tensor category (see \cite{HL5}).

Our approach is based on the formal calculus developed in \cite{FLM2},
and also (in later papers in this series) on the geometric methods
developed in \cite{H1}, \cite{H5}. Our use of formal calculus (see
\cite{FLM2},
\cite{FHL}) is equivalent to the use of contour integral methods, but
is far more natural and appropriate for our formulations and
arguments. For example, in Section 3, the space of rational functions
whose action we must define is described conceptually by means of the
formal $\delta$-function.

The formal-calculus techniques that we use in the present series of
papers might seem unfamiliar at first but we would like to reassure
the reader that these methods are as natural as any used in classical
mathematics, and are the right ones for the task.  A reader who spends
a bit of time to become familiar with the use of the basic properties
of the formal $\delta$-function will have little difficulty reading
the papers in this series.

Results in the present series of papers were announced in \cite{HL1}
and in talks presented by both authors at the June, 1992 AMS-IMS-SIAM
Joint Summer Research Conference on Conformal Field Theory,
Topological Field Theory and Quantum Groups at Mount Holyoke College.

A notion of tensor product related to but different {}from ours has
been studied in \cite{Li}. Also, related ideas, on a physical level of
rigor, based on suggestions of Borcherds, are discussed in
\cite{G}.

Part I is organized as follows: Section 2 reviews basic concepts in
the representation theory of vertex operator algebras. Section 3
discusses affinizations of vertex operator algebras, the opposite
module structure on a module for a vertex operator algebra and a
related $*$-operation. In \cite{B1}, in fact Borcherds placed a vertex
algebra structure on a certain affinization of a vertex algebra (in
his sense), while in this paper we are using more general
affinizations of a vertex operator algebra, but in a simpler way.
Section 4 gives the definitions of $P(z)$- and $Q(z)$-tensor product
of two modules for a vertex operator algebra and establishes some
straightforward consequences, including relations among intertwining
operators, ``intertwining maps'' and tensor products, and the
existence of a $Q(z)$-tensor product of the two modules for a rational
vertex operator algebra. In this section, we formulate and use a
result (Proposition 4.9) giving an isomorphism between certain spaces
of intertwining operators and we defer its proof to Part II
\cite{HL4}.  Sections 5 and 6 present the first and second
constructions of the $Q(z)$-tensor product of two modules,
respectively.  In the course of these constructions, we formulate and
use three results, Proposition 5.2, Theorem 6.1 and Proposition 6.2,
whose proofs will form the main content of Part II. Proposition 5.2 is
a commutator formula for vertex operators acting on the dual space of
the vector space $W_{1}\otimes W_{2}$, where $W_{1}$ and $W_{2}$ are
given modules whose vertex-operator-algebra-theoretic tensor product
we are studying. Theorem 6.1 states that the Jacobi identity holds
when we restrict to the subspace of $(W_{1}\otimes W_{2})^{*}$
consisting of the elements satisfying compatibility condition.
Proposition 6.2 asserts that this subspace is stable under the action
of our operators.  Theorem 6.1 and Proposition 6.2 are combined in
Theorem 6.3 to give a characterization of the tensor product of two
modules.  Theorem 6.1 and Proposition 6.2 and their extensions
constitute the foundation of the whole theory.

{\bf Historical note} \hspace{.1em}
Since the time that the present paper and Part
II \cite{HL4} were circulated as preprints starting in 1993 and were
submitted to the high-energy theoretical physics electronic preprint
archive {\tt hep-th} (as numbers 9309076 and 9309159, respectively),
we have expanded and updated the introduction and references and
corrected a few misprints; otherwise, Parts I and II are the same as
the original preprints.  Since then, the main technical part of the
theory, building on Theorem 6.3 of the present paper, has been carried
out, including the associativity.  See Part III \cite{HL6}, Part IV
\cite{H6} and the overview
\cite{HL5} for details. The theory has also been applied to the study of
the moonshine module in \cite{H7} and to the study of the
(nonmeromorphic) operator product expansion for intertwining operators
in \cite{H8}.

{\bf Acknowledgments}\hspace{.1em} We would like to thank D.~Kazhdan,
G.~Lusztig and M.~Finkelberg for interesting discussions, especially
concerning the comparison between their approach and ours in the case
of affine Lie algebras. We are also grateful to I.~M.~Gelfand for
initially directing our attention to the preprint of the paper
\cite{KL1} in his seminar at Rutgers University and to O.~Mathieu for
illuminating comments on that preprint. We thank R.~Borcherds for
informing us that some years ago, he also began considering a notion
of tensor product of modules for a vertex algebra.  During the course
of this work, Y.-Z.~H. was supported in part by NSF grants DMS-8610730
(through the Institute for Advanced Study), DMS-9104519 and
DMS-9301020 and J.~L. by NSF grants DMS-8603151 and DMS-9111945.

\renewcommand{\theequation}{\thesection.\arabic{equation}}
\renewcommand{\therema}{\thesection.\arabic{rema}}
\setcounter{equation}{0}

\section{Review of basic concepts}

In this section, we review some basic definitions and concepts in the
representation theory of vertex operator algebras. Except for
Definition 2.11, everything in this section can be found in
\cite{FLM2} and \cite{FHL}.

In this paper, all the variables $x$, $x_{0}, \dots$ are independent
commuting formal variables, and all expressions involving these
variables are to be understood as formal Laurent series or,
when explicitly so designated, as formal
rational functions. (Later, we shall also use the symbols $z,
z_{0}, \dots,$ which will denote complex numbers, {\it not} formal
variables.) We use the formal expansion
\begin{equation}
\delta(x)=\sum_{n\in
{\Bbb Z}}x^{n}.
\end{equation}
This ``formal $\delta$-function'' has the following simple and
fundamental property:
For any $f(x)\in {\Bbb C}[x, x^{-1}]$,
\begin{equation}
f(x)\delta(x)=f(1)\delta(x).
\end{equation}
This property has many important variants. For example, for any
$$X(x_{1},
x_{2})\in (\mbox{End }W)[[x_{1}, x_{1}^{-1}, x_{2}, x_{2}^{-1}]]$$
(where $W$ is a vector space) such that
\begin{equation}
\lim_{x_{1}\to x_{2}}X(x_{1}, x_{2})=X(x_{1},
x_{2})\lbar_{x_{1}=x_{2}}
\end{equation}
exists, we have
\begin{equation}
X(x_{1}, x_{2})\delta\left(\frac{x_{1}}{x_{2}}\right)=X(x_{2}, x_{2})
\delta\left(\frac{x_{1}}{x_{2}}\right).
\end{equation}
The existence of the ``algebraic limit'' defined in (2.3) means that
for an arbitrary vector $w\in W$, the coefficient of each power of
$x_{2}$ in the formal expansion $X(x_{1}, x_{2})w\lbar_{x_{1}=x_{2}}$
is a finite sum.  We use the convention that negative powers of a
binomial are to be expanded in nonnegative powers of the second
summand. For example,
\be
x_{0}^{-1}\delta\left(\frac{x_{1}-x_{2}}{x_{0}}\right)=\sum_{n\in {\Bbb Z}}
\frac{(x_{1}-x_{2})^{n}}{x_{0}^{n+1}}=\sum_{m\in {\Bbb N},\; n\in {\Bbb Z}}
(-1)^{m}{{n}\choose {m}} x_{0}^{-n-1}x_{1}^{n-m}x_{2}^{m}.
\ee
We have the following identities:
\begin{eqnarray}
&{\dps x_{1}^{-1}\delta\left(\frac{x_{2}+x_{0}}{x_{1}}\right)
=x_{2}^{-1}\left(
\frac{x_{1}-x_{0}}{x_{2}}\right),}&\\
&{\dps x_{0}^{-1}\delta\left(\frac{x_{1}-x_{2}}{x_{0}}\right)-
x_{0}^{-1}\delta\left(\frac{x_{2}-x_{1}}{-x_{0}}\right)=
x_{2}^{-1}\delta\left(\frac{x_{1}-x_{0}}{x_{2}}\right).}&
\end{eqnarray}
We shall use these properties and identities extensively later on
without explicit comment. See \cite{FLM2} and \cite{FHL} for further
discussion and many examples of their use.

We now quote the definition and basic ``duality''
properties of vertex operator algebras {from} \cite{FLM2} or \cite{FHL}:

\begin{defi}
{\rm A {\it vertex operator algebra} (over ${\Bbb C}$) is a ${\Bbb
Z}$-graded vector space (graded by {\it weights})
\begin{equation}
V=\coprod_{n\in {\Bbb Z}}V_{(n)}; \ \mbox{\rm for}\ v\in
V_{(n)},\;n=\mbox{\rm wt}\ v;
\end{equation}
such that
\begin{equation}
\mbox{\rm dim }V_{(n)}<\infty\;\;\mbox{\rm for}\; n \in {\Bbb Z},
\end{equation}
\begin{equation}
V_{(n)}=0\;\;\mbox{\rm for} \;n\; \mbox{\rm sufficiently small},
\end{equation}
equipped with a linear map  $V\otimes V\to V[[x, x^{-1}]]$, or
equivalently,
\begin{eqnarray}
V&\to&(\mbox{\rm End}\; V)[[x, x^{-1}]]\nonumber \\
v&\mapsto& Y(v, x)={\displaystyle \sum_{n\in{\Bbb Z}}}v_{n}x^{-n-1}
\;\;(\mbox{\rm where}\; v_{n}\in
\mbox{\rm End} \;V),
\end{eqnarray}
$Y(v, x)$ denoting the {\it vertex operator associated with} $v$, and
equipped also with two distinguished homogeneous vectors ${\bf 1}\in
V_{(0)}$ (the {\it vacuum}) and $\omega \in V_{(2)}$. The following
conditions are assumed for $u, v \in V$: the {\it lower truncation
condition} holds:
\begin{equation}
u_{n}v=0\;\;\mbox{\rm for}\;n\; \mbox{\rm sufficiently large}
\end{equation}
(or equivalently, $Y(u, x)v\in V((x))$);
\begin{equation}
Y({\bf 1}, x)=1\;\; (1\;\mbox{\rm on the right being the identity
operator});
\end{equation}
the {\it creation property} holds:
\begin{equation}
Y(v, x){\bf 1} \in V[[x]]\;\;\mbox{\rm and}\;\;\lim_{x\rightarrow
0}Y(v, x){\bf 1}=v
\end{equation}
(that is, $Y(v, x){\bf 1}$ involves only nonnegative integral powers
of $x$ and the constant term is $v$); the {\it Jacobi identity} (the
main axiom) holds:
\begin{eqnarray}
&x_{0}^{-1}\delta
\left({\displaystyle\frac{x_{1}-x_{2}}{x_{0}}}\right)Y(u, x_{1})Y(v,
x_{2})-x_{0}^{-1} \delta
\left({\displaystyle\frac{x_{2}-x_{1}}{-x_{0}}}\right)Y(v, x_{2})Y(u,
x_{1})&\nonumber \\ &=x_{2}^{-1} \delta
\left({\displaystyle\frac{x_{1}-x_{0}}{x_{2}}}\right)Y(Y(u, x_{0})v,
x_{2})&
\end{eqnarray}
(note that when each expression in (2.15) is applied to any element of
$V$, the coefficient of each monomial in the formal variables is a
finite sum; on the right-hand side, the notation $Y(\cdot, x_{2})$ is
understood to be extended in the obvious way to $V[[x_{0},
x^{-1}_{0}]]$); the Virasoro algebra relations hold:
\begin{equation}
[L(m), L(n)]=(m-n)L(m+n)+{\displaystyle\frac{1}{12}}
(m^{3}-m)\delta_{n+m,0}c
\end{equation}
for $m, n \in {\Bbb Z}$, where
\begin{equation}
L(n)=\omega _{n+1}\;\; \mbox{\rm for} \;n\in{\Bbb Z}, \;\;{\rm
i.e.},\;\;Y(\omega, x)=\sum_{n\in{\Bbb Z}}L(n)x^{-n-2}
\end{equation}
and
\begin{eqnarray}
&c\in {\Bbb C};&\\ &L(0)v=nv=(\mbox{\rm wt}\ v)v\;\;\mbox{\rm for}\;n
\in {\Bbb
Z}\;\mbox{\rm and}\;v\in V_{(n)};&\\ &{\displaystyle \frac{d}{dx}}Y(v,
x)=Y(L(-1)v, x)&
\end{eqnarray}
(the {\it  $L(-1)$-derivative property}).}
\end{defi}

The vertex operator algebra just defined is denoted by $(V, Y, {\bf 1},
\omega)$ (or simply by $V$). The complex number $c$ is called the {\it
central charge} or {\it rank} of $V$.  Homomorphisms of vertex
operator algebras are defined in the obvious way.

\begin{rema}
{\rm The axioms above imply that if $v\in V$ is
homogeneous and $n\in {\Bbb Z}$, \begin{equation} \mbox{\rm wt}\;
v_{n}=\mbox{\rm wt}\; v-n-1 \end{equation} as an operator. We shall
also use the fact that in the presence of other axioms, the Virasoro
algebra commutator relations (2.16) are equivalent to the relation
\begin{equation} Y(\omega ,x)\omega = {1\over 2} c {\bf 1}x^{-4} +
2\omega x^{-2} + L(-1)\omega x^{-1} + v \end{equation} where $v \in
V[[x]].$}
\end{rema}

Vertex operator algebras have important ``rationality,''
``commutativity'' and ``associativity'' properties, collectively
called ``duality'' properties.  These properties can in fact be used
as axioms replacing the Jacobi identity in the definition of vertex
operator algebra, as we now recall.

In the propositions below, ${\Bbb C}[x_{1}, x_{2}]_{S}$ is the ring of
rational functions obtained by inverting (localizing with respect to)
the products of (zero or more) elements of the set $S$ of nonzero
homogeneous linear polynomials in $x_{1}$ and $x_{2}$. Also,
$\iota_{12}$ (which might also be written as $\iota_{x_{1}x_{2}}$) is
the operation of expanding an element of ${\Bbb C}[x_{1}, x_{2}]_{S}$,
that is, a polynomial in $x_{1}$ and $x_{2}$ divided by a product of
homogeneous linear polynomials in $x_{1}$ and $x_{2}$, as a formal series
containing at most finitely many negative powers of $x_{2}$ (using
binomial expansions for negative powers of linear polynomials
involving both $x_{1}$ and $x_{2}$); similarly for $\iota_{21}$ and so
on. (The distinction between rational functions and formal Laurent
series is crucial.)

For any ${\Bbb Z}$-graded, or more generally, ${\Bbb C}$-graded, vector space
$W=\coprod W_{(n)}$, we use the notation
\begin{equation}
W'=\coprod W_{(n)}^{*}
\end{equation}
for its graded dual.
\begin{propo}
{\bf (a) (rationality of products)} For $v$, $v_{1}$, $v_{2}\in V$ and
$v'\in V'$, the formal series
$\left\langle v', Y(v_{1}, x_{1})Y(v_{2}, x_{2})v\right\rangle,$ which
involves only finitely many negative powers of $x_{2}$ and only
finitely many positive powers of $x_{1}$, lies in the image of the map
$\iota_{12}$:
\begin{equation}
\left\langle v', Y(v_{1}, x_{1})Y(v_{2}, x_{2})v\right\rangle
=\iota_{12}f(x_{1}, x_{2}),
\end{equation}
where the (uniquely determined) element $f\in {\Bbb C}[x_{1},
x_{2}]_{S}$ is of the form
\begin{equation}
f(x_{1}, x_{2})={\displaystyle \frac{g(x_{1},
x_{2})}{x_{1}^{r}x_{2}^{s}(x_{1}-x_{2})^{t}}}
\end{equation}
for some $g\in {\Bbb C}[x_{1}, x_{2}]$ and $r, s, t\in {\Bbb Z}$.

{\bf (b) (commutativity)} We also have
\begin{equation}
\left\langle v', Y(v_{2}, x_{2})Y(v_{1}, x_{1})v\right\rangle
=\iota_{21}f(x_{1}, x_{2}). \hspace{3em}\Box
\end{equation}
\end{propo}
\begin{propo}
{\bf (a) (rationality of iterates)} For $v$, $v_{1}$, $v_{2}\in V$ and
$v'\in V'$, the formal series
$\left\langle v', Y(Y(v_{1}, x_{0})v_{2}, x_{2})v\right\rangle,$
which involves only finitely many negative powers of $x_{0}$ and only
finitely many positive powers of $x_{2}$, lies in the image of the map
$\iota_{20}$:
\begin{equation}
\left\langle v', Y(Y(v_{1}, x_{0})v_{2},
x_{2})v\right\rangle=\iota_{20}h(x_{0}, x_{2}),
\end{equation}
where the (uniquely determined) element $h\in {\Bbb C}[x_{0},
x_{2}]_{S}$ is of the form
\begin{equation}
h(x_{0}, x_{2})={\displaystyle \frac{k(x_{0},
x_{2})}{x_{0}^{r}x_{2}^{s}(x_{0}+x_{2})^{t}}}
\end{equation}
for some $k\in {\Bbb C}[x_{0}, x_{2}]$ and $r, s, t\in {\Bbb Z}$.

{\bf (b)} The formal series
$\left\langle v', Y(v_{1}, x_{0}+x_{2})Y(v_{2}, x_{2})v\right\rangle,$
which involves only finitely many negative powers of $x_{2}$ and only
finitely many positive powers of $x_{0}$, lies in the image of
$\iota_{02}$, and in fact
\begin{equation}
\left\langle v', Y(v_{1}, x_{0}+x_{2})Y(v_{2},
x_{2})v\right\rangle=\iota_{02}h(x_{0}, x_{2}).\hspace{3em}\Box
\end{equation}
\end{propo}
\begin{propo}[associativity]
We have the following equality of rational functions:
\begin{equation}
\iota_{12}^{-1}\left\langle v', Y(v_{1}, x_{1})Y(v_{2}, x_{2})v\right\rangle
=(\iota_{20}^{-1}\left\langle v', Y(Y(v_{1}, x_{0})v_{2},
x_{2})v\right\rangle)\lbar_{x_{0}=x_{1}-x_{2}}.\hspace{1.5em}\Box
\end{equation}
\end{propo}
\begin{propo}
In the presence of the other axioms, the Jacobi identity follows {from}
the rationality of products and iterates, and commutativity and
associativity. In particular, in the definition of vertex operator
algebra, the Jacobi identity may be replaced by these properties.\epf
\end{propo}

We have the following notions of module and of intertwining operator for
vertex operator algebras:

\begin{defi}
{\rm  Given a vertex operator algebra
$(V,Y,{\bf 1},\omega )$,  a {\it module for  $V$}
(or  $V$-{\it module}  or {\it representation space}) is
a  ${\Bbb C}$-graded  vector space (graded by {\it weights})
\begin{equation}
W =\coprod_{n\in {\Bbb C}}W_{(n)}; \;\;\mbox{for}\;\;w \in  W_{(n)},
\;\; n = \mbox{wt}\ w;
\end{equation}
 such that
\begin{equation}
\dim \ W_{(n)} < \infty
\;\;\mbox{for}\;\;n \in {\Bbb C},
\end{equation}
\begin{equation}
W_{(n)} = 0 \;\; \mbox{for}\;\; n \;\; \mbox{whose real part is
sufficiently small,}
\end{equation}
equipped with a linear map
$V\otimes W \rightarrow W[[x,x^{-1}]]$, or equivalently,
\begin{eqnarray}
V &\rightarrow & (\mbox{End}\ W)[[x,x^{-1}]]\nno\\ v
&\mapsto & Y(v,x) =\sum_{n\in {\Bbb Z}}v_nx^{-n-1}\;\;\;
(\mbox{where}\;\; v_n \in \mbox{End}\ W)
\end{eqnarray}
(note that the
sum is over ${\Bbb Z}$, not ${\Bbb C}$), $Y(v,x)$ denoting the {\it
vertex operator associated with $v$}, such that ``all the defining
properties of a vertex operator algebra that make sense hold."  That
is, for $u,v \in V$ and $w \in W$,
\begin{equation}
v_nw = 0 \;\;
\mbox{for} \;\;n \;\; \mbox{sufficiently large}
\end{equation}
(the
lower truncation condition);
\begin{equation}
Y(\mbox{\bf 1},z) = 1;
\end{equation}
\begin{eqnarray}
&{\dps x^{-1}_0\delta \left(
{x_1-x_2\over x_0}\right) Y(u,x_1)Y(v,x_2) - x^{-1}_0\delta \left(
{x_2-x_1\over -x_0}\right) Y(v,x_2)Y(u,x_1)}&\nno\\ &{\dps =
x^{-1}_2\delta \left( {x_1-x_0\over x_2}\right) Y(Y(u,x_0)v,x_2)}
\end{eqnarray}
(the Jacobi identity for operators on $W)$; note that
on the right-hand side, $Y(u,x_0)$ is the operator associated with
$V$; the Virasoro algebra relations hold on $W$ with scalar $c$ equal
to the central charge of $V$:
\begin{equation}
[L(m),L(n)] =
(m-n)L(m+n) + {1\over 12}(m^3-m)\delta _{m+n,0}c
\end{equation} for
$m,n \in {\Bbb Z}$, where
\begin{equation}
L(n) = \omega _{n+1}\;\;
\mbox{for}\;\; n \in {\Bbb Z}, \;\; \mbox{i.e.}, \;\; Y(\omega ,z)
=\sum_{n\in {\Bbb Z}}L(n)x^{-n-2};
\end{equation}

\begin{equation}
L(0)w = nw = (\mbox{wt}\ w)w \;\; \mbox{for}\;\; n \in  {\Bbb C}
\;\;\mbox{and}\;\;  w \in  W_{(n)};
\end{equation}
\begin{equation}
{d\over dx}Y(v,x) = Y(L(-1)v,x),
\end{equation}
where  $L(-1)$  is the operator on  $V$. }
\end{defi}

This completes the definition of module.
We may denote the module just defined by
$(W,Y)$ (or simply by $W$).
If necessary, we shall use $Y_{W}$ or similar notation to indicate
that the vertex operators concerned act on $W$. Homomorphisms (or
maps) of
$V$-modules  are defined in the obvious way. For $V$-modules $W_{1}$
and $W_{2}$, we shall denote the space of module maps {from} $W_{1}$ to
$W_{2}$ by Hom$_{V}(W_{1}, W_{2})$.

\begin{rema}
{\rm Formula (2.21) holds for modules. Also note that
the Virasoro algebra commutator relations (2.38) are in
fact  consequences of the other axioms, in view of (2.22).}
\end{rema}

For any vector space $W$ and any formal variable $x$, we use the notation
\begin{equation}
W\{ x\}=\biggl\{\sum_{n\in {\Bbb C}}a_{n}x^{n}|
a_{n}\in W,\; n\in {\Bbb C}\biggr\}.
\end{equation}
In particular, we shall allow complex powers of our commuting formal
variables.
\begin{defi}
{\rm Let  $V$  be a vertex operator algebra and let
$(W_1,Y_1)$,  $(W_2,Y_2)$  and  $(W_3,Y_3)$  be three  $V$-modules  (not
necessarily distinct, and possibly equal to  $V)$.  An {\it intertwining
operator of  type
${W_3}\choose {W_1\ W_2} $}  is a linear map  $W_1\otimes W_2
\rightarrow  W_3\{x\}$,  or equivalently,
\begin{eqnarray}
W_1 &\rightarrow & (\mbox{Hom}(W_2,W_3))\{x\}\nno\\
w & \mapsto & {\cal Y}(w,x) =\sum_{n\in
{\Bbb C}}w_nx^{-n-1}\;\;\; (\mbox{where}\;\;  w_n \in  \mbox{Hom}(W_2,W_3))
\end{eqnarray}
such that ``all the defining properties of a module action that make sense
hold."  That is, for  $v \in  V$,  $w_{(1)} \in  W_1$ and
$w_{(2)} \in  W_2,$ we have the lower truncation condition
\begin{equation}
(w_{(1)})_{n}w_{(2)} = 0\;\;  \mbox{for}\;\; n \;\; \mbox{whose real part is
sufficiently large;}
\end{equation}
the following Jacobi identity holds for the operators  $Y_{1}(v,\cdot )$,
$Y_{2}(v,\cdot )$, $Y_{3}(v,\cdot )$ and
${\cal Y}(\cdot, x_{2})$  acting on the element  $w_{(2)}$:
\begin{eqnarray}
\lefteqn{\dps x^{-1}_0\delta \left( {x_1-x_2\over x_0}\right)
Y_3(v,x_1){\cal Y}(w_{(1)},x_2)w_{(2)}}\nno\\
&&\hspace{2em}- x^{-1}_0\delta \left( {x_2-x_1\over -x_0}\right)
{\cal Y}(w_{(1)},x_2)Y_2(v,x_1)w_{(2)}\nno \\
&&{\dps = x^{-1}_2\delta \left( {x_1-x_0\over x_2}\right)
{\cal Y}(Y_1(v,x_0)w_{(1)},x_2)
w_{(2)}}
\end{eqnarray}
(note that the first term on the left-hand side is algebraically meaningful
because of the condition (2.44), and the other terms are
meaningful by the usual
properties of modules; also note that this Jacobi identity involves integral
powers of  $x_0$ and  $x_1$ and complex powers of  $x_2$);
\begin{equation}
{d\over dx}{\cal Y}(w_{(1)},x) = {\cal Y}(L(-1)w_{(1)},x),
\end{equation}
where  $L(-1)$  is the operator acting on  $W_{1}$. }
\end{defi}

The intertwining operators of the
same type ${W_{3}}\choose {W_{1}\ W_{2}}$ form a vector space, which we
denote  by ${\cal
V}^{W_{3}}_{W_{1}W_{2}}$. The dimension of this vector space
is called the {\it fusion rule} for $W_{1}$,
$W_{2}$ and $W_{3}$ and is denoted by
$N^{W_{3}}_{W_{1}W_{2}}$ ($\le \infty$).  Formula (2.21) holds for intertwining
operators,
with $v_{n}$ replaced by $w_{n}$ ($n\in {\Bbb C}$).

There are also duality properties for modules and intertwining operators. See
\cite{FHL} and \cite{DL} for details.

Let  $(W,Y)$,  with
\begin{equation}
W = \coprod_{n\in {\Bbb C}}W_{(n)},
\end{equation}
be a module for a vertex operator algebra  $(V,Y,{\bf 1},\omega ),$
and consider its graded dual space
$W'$ (recall (2.23)).  We define the {\it contragredient
vertex operators} (called ``adjoint vertex operators'' in \cite{FHL})
  $Y'(v,x)$ $ (v \in  V)$ by means of the linear map
\begin{eqnarray}
V &\rightarrow&  (\mbox{End}\ W')[[x,x^{-1}]]\nno\\
v &\mapsto & Y'(v,x) = \sum_{n\in
{\Bbb Z}}v'_nx^{-n-1} \;\;\;(\mbox{where}  \;\;v'_n \in  \mbox{End}\ W'),
\end{eqnarray}
determined by the condition
\begin{equation}
\langle Y'(v,x)w',w\rangle  =
\langle w',Y(e^{xL(1)}(-x^{-2})^{L(0)}v,x^{-1})w\rangle
\end{equation}
for  $v \in  V$,  $w' \in  W^{'}$,  $w \in  W$.
The operator
$(-x^{-2})^{L(0)}$ has the obvious meaning; it acts on a vector of weight  $n
\in  {\Bbb Z}$  as multiplication by  $(-x^{-2})^n$.  Also note that
$e^{xL(1)}(-x^{-2})^{L(0)}v$  involves only finitely many (integral) powers of
$x$,  that the right-hand side of (2.49) is a Laurent polynomial in  $x$,  and
that  the components  $v'_n$  of the formal Laurent series
$Y'(v,x)$ indeed preserve  $W'.$

We give the space  $W'$ a  ${\Bbb C}$-grading by setting
\begin{equation}
W'_{(n)} = W_{(n)}^{*}\;\; \mbox{for}\;\;  n \in  {\Bbb C}
\end{equation}
(cf. (2.23)). The following theorem defines the  $V$-{\it module $W'$
contragredient to  $W$}
(see \cite{FHL}, Theorem 5.2.1 and Proposition 5.3.1):
\begin{theo}
 The pair  $(W',Y')$  carries the structure of a
$V$-module and $(W'', Y'')=(W, Y)$.\epf
\end{theo}

Given a module map $\eta: W_{1}\to W_{2}$, there is a unique module
map $\eta': W'_{2}\to W'_{1}$,  the {\it adjoint} map, such
that
\begin{equation}
\langle \eta'(w'_{(2)}), w_{(1)}\rangle=\langle w'_{(2)},
\eta(w_{(1)}\rangle
\end{equation}
for $w_{(1)}\in W_{1}$ and $w_{(2)}\in W_{2}$. (Here the pairings
$\langle \cdot, \cdot \rangle$ on the two sides refer to two
different modules.) Note that
\begin{equation}
\eta''=\eta.
\end{equation}

In the construction of the tensor product module
of two modules for a vertex operator
algebra, we shall need the following generalization of
the notion of module recalled above:
\begin{defi}
{\rm A
{\it  generalized $V$-module} is a
${\Bbb C}$-graded vector space $W$  equipped with a linear map of the form
(2.34) satisfying all the axioms for a  $V$-module except
that
 the homogeneous subspaces need not be finite-dimensional
and
that they need not be zero even for $n$ whose real part is
sufficiently small; that is, we omit (2.32) and (2.33) {from} the definition. }
\end{defi}

\renewcommand{\theequation}{\thesection.\arabic{equation}}
\renewcommand{\therema}{\thesection.\arabic{rema}}
\setcounter{equation}{0}
\setcounter{rema}{0}

\section{Affinizations of vertex operator algebras and the $*$-operation}

In order to use the Jacobi identity to construct a tensor product of
modules for a vertex operator algebra, we shall study  various
``affinizations'' of a vertex operator algebra with respect to certain
algebras and vector spaces of formal Laurent series and formal
rational functions.

Let
$V$ be a vertex operator algebra and $W$ a $V$-module. We adjoin the
formal variable $t$ to our list of commuting formal variables. This
variable will play a special role.
Consider the vector spaces $$V[t,t^{-1}] = V \otimes {\Bbb C}[t,t^{-1}]
\subset V \otimes {\Bbb C} ((t)) \subset V\otimes {\Bbb C} [[t,t^{-1}]]
\subset V[[t,t^{-1}]] $$ (note carefully the distinction between the
last two, since $V$ is typically infinite-dimensional) and $W \otimes
{\Bbb C} \{t\}
\subset W \{t\}$
(recall (2.42)). The linear map
\begin{eqnarray}
\tau_W: V[t,t^{-1}] &\to& \mbox{End} \;W\nno\\
v \otimes t^n &\mapsto&
v_n
\end{eqnarray}
($v\in V$, $n\in {\Bbb Z}$) extends canonically to
\begin{eqnarray}
\tau_W:& V \otimes {\Bbb
C}((t)) &\to\; \mbox{End} \;W \nno\\
&v \otimes {\dps \sum_{n > N}} a_nt^n
&\mapsto\; \sum_{n > N} a_nv_n
\end{eqnarray}
 (but not to $V((t))$), in view of (2.21).
It further extends canonically to
\begin{equation}
\tau_W: (V
\otimes {\Bbb C}((t)))[[x,x^{-1}]] \to (\mbox{End} \;W)[[x,x^{-1}]],
\end{equation}
where of course $(V
\otimes {\Bbb C}((t)))[[x,x^{-1}]]$ can be viewed as the subspace of
$V[[t,t^{-1},x,x^{-1}]]$ such that the coefficient of each power of
$x$ lies in $V \otimes {\Bbb C}((t))$.

Let $v \in V$ and define the ``generic vertex operator''
\begin{equation}
Y_t(v,x)  = \sum_{n \in {\Bbb Z}}(v \otimes
t^n)x^{-n-1} \in  (V \otimes {\Bbb C}[t,t^{-1}])[[x,x^{-1}]].
\end{equation}
Then
\begin{eqnarray}
Y_t(v,x) & = & v\otimes x^{-1} \delta \left(\frac{t}{x}\right)\nno\\
 & = & v \otimes t^{-1} \delta \left(\frac{x}{t}\right)\nno\\
 & \in & V \otimes {\Bbb C}
[[t,t^{-1},x,x^{-1}]]\nno\\
 & (\subset & V[[t,t^{-1},x,x^{-1}]])
\end{eqnarray}
and the linear map
\begin{eqnarray}
V& \to &V\otimes {\Bbb C}[[t,t^{-1},x,x^{-1}]]\nno\\
v &\mapsto & Y_t(v,x)
\end{eqnarray}
 is simply the  map  given by tensoring by the
``universal element'' $x^{-1} \delta
\left(\frac{t}{x}\right)$.  We have
\begin{equation}
\tau_W(Y_t(v,x)) =
Y_W(v,x).
\end{equation}
 For all $f(x) \in {\Bbb C} [[x,x^{-1}]]$,
$f(x)Y_t(v,x)$ is defined and
\begin{equation}
f(x)Y_t(v,x) = f(t) Y_t(v,x).
\end{equation}
In case $f(x) \in {\Bbb C}((x))$, then $\tau_{W}(f(x)Y_{t}(v, x))$ is
also defined, and
\begin{equation}
f(x)Y_W(v,x)  =  f(x)\tau_W(Y_t(v,x)) =
\tau_W(f(x)Y_t(v,x)) =  \tau_W(f(t)Y_t(v,x)).
\end{equation}
The expansion coefficients, in powers of $x$, of $Y_t(v,x)$
span $v \otimes {\Bbb C}[t,t^{-1}],$ the $x$-expansion coefficients
of $Y_W(v,x)$ span $\tau_W(v \otimes {\Bbb C}[t,t^{-1}])$ and the
$x$-expansion coefficients of $f(x)Y_t(v,x)$ span $v \otimes f(t)
{\Bbb C} [t,t^{-1}]$. In case $f(x) \in {\Bbb C}((x))$, the
$x$-expansion coefficients of $f(x)Y_W(v,x)$ span $\tau_W(v \otimes
f(t) {\Bbb C}[t,t^{-1}])$.

Using this viewpoint, we shall examine each of the three terms in
 the Jacobi identity (2.45) in the definition of
intertwining operator. First we consider the formal Laurent series in
$x_{0}$, $x_{1}$, $x_{2}$ and $t$ given by
\begin{eqnarray}
&{\dps x^{-1}_2 \delta\left(\frac{x_1-x_0}{x_2}\right)
Y_t(v,x_0) = x^{-1}_1\delta \left(\frac{x_2+x_0}{x_1}\right)
Y_t(v,x_0)}&\nno\\
&{\dps = v \otimes x^{-1}_1\delta\left(\frac{x_2 +
t}{x_1}\right)x^{-1}_0 \delta\left(\frac{t}{x_0}\right)}&
\end{eqnarray}
(cf. the right-hand side of (2.45)). The
expansion coefficients in powers of $x_0$, $x_1$ and $x_2$ of (3.10)
span just the space $v \otimes {\Bbb C}[t,t^{-1}]$. However,
the expansion coefficients in $x_0$ and $x_1$ only (but not in $x_{2}$) of
\begin{eqnarray}
\lefteqn{x^{-1}_1\delta\left(\frac{x_{2}+x_0}{x_1}\right) Y_t(v,x_0)
=} \nno\\
& & = v \otimes x^{-
1}_1\delta\left(\frac{x_{2}+t}{x_1}\right)x^{-
1}_0\delta\left(\frac{t}{x_0}\right)\nno\\
& & = v \otimes
\left(\sum_{m \in {\Bbb Z}} (x_{2} + t)^m x^{-m-
1}_1\right)\left(\sum_{n \in {\Bbb Z}}t^n x^{-n-
1}_0\right)
\end{eqnarray}
 span $$v \otimes \iota_{x_{2},t}{\Bbb C}[t,t^{- 1}, x_{2}+t,
(x_{2}+t)^{-1}]\subset v \otimes {\Bbb C}[x_{2}, x_{2}^{-1}]((t)),$$
where $\iota_{x_{2}, t}$ is the operation of expanding a formal rational
function in the indicated algebra as a formal Laurent series
involving only finitely many negative powers of $t$ (cf.
the notation $\iota_{12}$, etc., above). We shall use similar
$\iota$-notations below. Specifically, the coefficient of
$x^{-n-1}_0 x^{-m-1}_1$ $(m,n
\in {\Bbb Z})$ in (3.11) is $v \otimes (x_{2} + t)^m t^n$.

We may
specialize $x_2 \mapsto z\in {\Bbb C}^{\times}$, and (3.11) becomes
\begin{eqnarray}
\lefteqn{z^{-1}\delta\left(\frac{x_{1}-x_0}{z}\right) Y_t(v,x_0)
=} \nno\\
&&=x^{-1}_1\delta\left(\frac{z+x_0}{x_1}\right) Y_t(v,x_0)\nno\\
& & = v \otimes x^{-
1}_1\delta\left(\frac{z+t}{x_1}\right)x^{-
1}_0\delta\left(\frac{t}{x_0}\right)\nno\\
& & = v \otimes
\left(\sum_{m \in {\Bbb Z}} (z + t)^m x^{-m-
1}_1\right)\left(\sum_{n \in {\Bbb Z}}t^n x^{-n-
1}_0\right).
\end{eqnarray}
The coefficient of $x^{-n-1}_0 x^{-m-1}_1$ $(m,n \in {\Bbb
Z})$ in (3.12) is $v \otimes (z + t)^m t^n\in V\otimes {\Bbb C}((t))$, and
these coefficients span
\begin{equation}
v\otimes {\Bbb C}[t, t^{-1}, (z+t)^{-1}]\subset v\otimes {\Bbb C}((t)).
\end{equation}
Our tensor product construction will be based on a certain action of the
space $V\otimes {\Bbb C}[t, t^{-1}, (z+t)^{-1}]$, and the description of
this space as the span of the coefficients of the expression (3.12) (as
$v\in V$ varies) will be very useful.

Now consider
\begin{eqnarray}
\lefteqn{x^{-1}_0 \delta\left(\frac{-x_2 +
x_1}{x_0}\right) Y_t(v,x_1)=}\nno\\
&& =  v
\otimes x^{-1}_0 \delta\left(\frac{-x_2 + t}{x_0}\right) x^{-1}_1
\delta \left(\frac{t}{x_1}\right)\nno\\
&&{\dps  =  v\otimes \biggr(\sum_{n \in {\Bbb Z}} (-x_2 + t)^n x^{-n-
1}_0\biggr)\biggr(\sum_{m \in {\Bbb Z}} t^m x^{-m-1}_1\biggr)}
\end{eqnarray}
(cf. the second term on the left-hand side of (2.45)).
The expansion coefficients in powers of $x_0$ and $x_1$ (but not $x_2$)
span
$$v \otimes \iota_{x_{2}, t}{\Bbb C}[t,t^{-1},-x_2 + t, (-x_2 + t)^{-1}],$$
and in fact the coefficient of $x^{-n-1}_0 x^{-m-1}_1$ $(m,n \in {\Bbb
Z})$ in (3.14) is $v
\otimes (-x_2 + t)^n t^m$. Again specializing $x_2 \mapsto z\in
{\Bbb C}^{\times}$, we obtain
\begin{eqnarray}
\lefteqn{x^{-1}_0 \delta \left(\frac{-z+x_1}{x_0}\right)
Y_t(v,x_1) =}\nno\\ &&= v\otimes x^{-1}_0 \delta \left(\frac{-
z+t}{x_0}\right) x^{-1}_1 \delta \left(\frac{t}{x_1}\right) \nno\\ & &
= v \otimes \biggr(\sum_{n \in {\Bbb Z}} (-z+t)^n x^{-n- 1}_0\biggr)
\biggr(\sum_{m \in
{\Bbb Z}}t^m x^{-m-1}_1\biggr).
\end{eqnarray}
 The coefficient of $x^{-n-1}_0 x^{-
m-1}_1$ $(m,n \in {\Bbb
Z})$ in (3.15)   is $v \otimes (-z+t)^n t^m$, and these coefficients span
\begin{equation}
v\otimes {\Bbb C}[t, t^{-1}, (-z+t)^{-1}]\subset v\otimes {\Bbb C}((t)).
\end{equation}

Finally, consider
\begin{eqnarray}
\lefteqn{\dps x^{-1}_0 \delta \left(\frac{x_1 -
x_2}{x_0}\right) Y_t (v,x_1)=}\nno\\
&&{\dps = v \otimes x^{-1}_0
\delta \left(\frac{t-x_2}{x_0}\right) x^{-1}_1 \delta
\left(\frac{t}{x_1}\right).}
\end{eqnarray}
The coefficient of $x_{0}^{-n-1}x_{1}^{-m-1}$ ($m, n\in {\Bbb Z}$) is
$v\otimes (t-x_{2})^{n}t^{m}$, and these expansion cofficients  span
$$v\otimes \iota_{t, x_{2}}{\Bbb C}[t, t^{-1}, t-x_{2}, (t-x_{2})^{-1}].$$
If we again specialize $x_2
\mapsto z$, we get
\begin{equation}
x^{-1}_0\delta\left(\frac{x_1-z}{x_0}\right)Y_t(v,x_1) = v\otimes
x_{0}^{-1}\delta\left(\frac{t-z}{x_{0}}\right)x_{1}^{-1}\delta\left(
\frac{t}{x_{1}}\right),
\end{equation}
whose coefficient of $x^{-n-1}_0 x^{-m-1}_1$  is $v
\otimes (t - z)^n t^m$.  These coefficients span
\begin{equation}
v\otimes {\Bbb C}[t, t^{-1}, (t-z)^{-1}]\subset v\otimes {\Bbb
C}((t^{-1}))
\end{equation}
(cf. (3.13), (3.16)).

Later we shall evaluate the identity (2.45) on the elements of the
contragredient module $W'_{3}$. This will allow us to convert the
expansion (3.19) into an expansion in positive powers of $t$.
It will be useful to examine the
notion of contragredient vertex operator ((2.48), (2.49)) more
closely. For a $V$-module $W$, we define the {\it opposite vertex
operator} associated to $v\in V$ by
\begin{equation}
Y^*_W(v,x) = Y_W(e^{xL(1)}(-
x^{-2})^{L(0)}v,x^{-1})
\end{equation}
 and we define its components by:
\begin{equation}
Y^*_W(v,x) = \sum_{n \in {\Bbb Z}} v^*_n x^{-n-1}.
\end{equation}
Then $v^*_n \in \mbox{End}\; W$ and $v \mapsto Y^*_W(v,x)$ is a linear
map $V \to (\mbox{End} \ W)[[x,x^{-1}]]$ such that $V \otimes W
\to W((x^{-1}))$ ($v \otimes w \mapsto Y^*_W(v,x)w).$ Note that the
contragredient vertex operators are the adjoints of the opposite
vertex operators:
\begin{equation}
\langle w', Y^{*}_{W}(v, x)w\rangle=\langle Y'(v, x)w', w\rangle
\end{equation}
and that if $v$ is homogeneous, the weight of the operator $v_{n}^{*}$
is $n+1-\mbox{\rm wt}\;v$, by (2.21).
The proof of Theorem 5.2.1 in \cite{FHL}, which asserts that $(W',
Y')$ is a $V$-module, in fact
proves the following {\it opposite Jacobi identity} for $Y^{*}_{W}$:
\bea
\lefteqn{\dps x_{0}^{-1}\delta\left(\frac{x_{1}-x_{2}}{x_{0}}\right)
Y_{W}^{*}(v_{2}, x_{2})Y^{*}_{W}(v_{1}, x_{1})}\nno\\
&&\hspace{2em}-x_{0}^{-1}\delta\left(\frac{x_{2}-x_{1}}{-x_{0}}\right)
Y_{W}^{*}(v_{1}, x_{1})Y^{*}_{W}(v_{2}, x_{2})\nno\\
&&{\dps =x_{2}^{-1}\delta\left(\frac{x_{1}-x_{0}}{x_{2}}\right)
Y_{W}^{*}(Y(v_{1}, x_{0})v_{2}, x_{2})}
\eea
(which of course also follows {from} the assertion that $(W', Y')$ is a
$V$-module).  The pair $(W, Y^{*})$ should be thought of as a ``right
module" for $V$.

We shall interpret the operator $Y^{*}_{W}$ by means of a $*$-operation
on $V\otimes {\Bbb C}[[t, t^{-1}]]$. This operation will be an involution.
We proceed as follows: First we generalize $Y^{*}$ in the following way:
Given any vector space $U$ and any linear map
\begin{eqnarray}
Z(\cdot,x):\ V & \rightarrow &  U[[x,x^{-
1}]]\;\;\biggr(=\prod_{n \in {\Bbb Z}} U \otimes x^n\biggr)\nno\\
v & \mapsto & Z(v,x)
\end{eqnarray}
{from} $V$ into $U[[x,x^{-1}]]$
(i.e., given any family of linear maps {from} $V$ into the spaces
$U \otimes x^n$), we define $Z^*(\cdot,x):\ V \to U[[x,x^{-1}]]$
by
\begin{equation}
Z^*(v,x) = Z(e^{xL(1)}(-x^{-2})^{L(0)}v,x^{-1}),
\end{equation}
 where
we use the obvious linear map $Z(\cdot,x^{-1}):\ V \to U[[x,x^{-
1}]],$ and where we extend $Z(\cdot,x^{-1})$ canonically to a
linear map $Z(\cdot,x^{-1}):\ V[x,x^{-1}] \to U[[x,x^{-1}]].$
Then by formula (5.3.1) in \cite{FHL}  (the proof of Proposition
5.3.1), we have
\begin{eqnarray}
Z^{**}(v,x) & = & Z^*(e^{xL(1)}(-x^{-
2})^{L(0)}v,x^{-1})\nno\\
& = & Z(e^{x^{-1}L(1)}(-
x^2)^{L(0)}e^{xL(1)}(-x^{-2})^{L(0)}v,x)\nno\\
&=& Z(v,x).
\end{eqnarray}
That is,
\begin{equation}
Z^{**}(\cdot,x) = Z(\cdot,x).
\end{equation}
 Moreover, if
$Z(v,x) \in U((x))$, then $Z^*(v,x) \in U((x^{-1}))$ and vice
versa.

     Now we expand $Z(v,x)$ and $Z^*(v,x)$ in components. Write
\begin{equation}
Z(v,z) = \sum_{n \in {\Bbb Z}}v_{(n)} x^{-n-1},
\end{equation}
 where for all $n\in {\Bbb Z}$,
\begin{eqnarray}
V & \to & U\nno\\
v & \mapsto & v_{(n)}
\end{eqnarray}
 is a linear map depending on $Z(\cdot, x)$ (and in fact, as
$Z(\cdot, x)$ varies, these linear maps are arbitrary).
Also write
\begin{equation}
Z^*(v,x)
= \sum_{n \in {\Bbb Z}} v^*_{(n)}x^{-n-1}
\end{equation}
 where
\begin{eqnarray}
V & \to & U\nno\\
v & \mapsto & v^*_{(n)}
\end{eqnarray}
 is a linear map depending on $Z(\cdot, x)$.  We shall compute $v_{(n)}^{*}$.
First note that
\begin{equation}
\sum_{n
\in {\Bbb Z}}v^*_{(n)}x^{-n-1} = \sum_{n \in {\Bbb Z}}(e^{xL(1)}(-
x^{-2})^{L(0)}v)_{(n)}x^{n+1}.
\end{equation}
  For convenience, suppose that $v
\in V_{(h)}$, for $h \in {\Bbb Z}$.  Then the right-hand side of (3.32)
is equal to
\begin{eqnarray}
\lefteqn{(-1)^h\sum_{n \in {\Bbb Z}}(e^{xL(1)}v)_{(-
n)}x^{-n+1-2h}}\nno\\
&&=  (-1)^h \sum_{n \in {\Bbb Z}} \sum_{m \in
{\Bbb N}} \frac{1}{m!}(L(1)^m v)_{(-n)}x^{m-n+1-2h}\nno\\
&&=
(-1)^h\sum_{m
\in {\Bbb N}} \frac{1}{m!} \sum_{n \in {\Bbb Z}}(L(1)^mv)_{(-n-m-
2+2h)}x^{-n-1},
\end{eqnarray}
that is,
\begin{equation}
v^*_{(n)} = (-
1)^h\sum_{m \in {\Bbb N}}\frac{1}{m!} (L(1)^mv)_{(-n-m-2+2h)}.
\end{equation}
For $v\in V$ not necessarily homogeneous, $v^*_{(n)}$ is given by the
appropriate sum of such expressions.

     Now consider the special case where $U = V \otimes {\Bbb
C}[t,t^{-1}]$ and where $Z(\cdot, x)$ is the ``generic'' linear map
\begin{eqnarray}
Y_t(\cdot,x): V & \rightarrow & (V \otimes
{\Bbb C}[t,t^{-1}])[[x,x^{-1}]]\nno\\
v & \mapsto & Y_t(v,x) = \sum_{n
\in {\Bbb Z}}(v \otimes t^n)x^{-n-1}
\end{eqnarray}
(recall (3.4)), i.e.,
\begin{equation}
v_{(n)} = v \otimes t^n.
\end{equation}
Then for $v\in V_{(h)}$,
\begin{equation}
v^*_{(n)} = (-1)^h\sum_{m \in {\Bbb N}}
\frac{1}{m!}((L(1))^mv) \otimes t^{-n-m-2+2h}
\end{equation}
 in this case.

 This motivates defining a $*$-operation on $V \otimes {\Bbb
C}[t,t^{-1}]$ as follows: For any $n, h\in {\Bbb Z}$ and $v\in V_{(h)}$,
define
\begin{equation}
(v\otimes t^n)^* = (-1)^h\sum_{m \in {\Bbb N}} \frac{1}{m!}
(L(1)^mv) \otimes t^{-n-m-2+2h}\in V \otimes {\Bbb C}[t,t^{-
1}],
\end{equation}
and extend by linearity to $V \otimes {\Bbb C}[t,t^{-1}]$.
That is, $(v \otimes t^n)^* = v^*_{(n)}$ for the special case $Z(\cdot, x)
= Y_t(\cdot, x)$ discussed above. (Note that for general $Z$,
we cannot expect to
be able to define an analogous $*$-operation on $U$.)  Also consider the map
\begin{eqnarray}
Y^*_t(\cdot,x) =
(Y_t(\cdot,x))^*:  V & \rightarrow & (V \otimes {\Bbb C}[t,t^{-
1}])[[x,x^{-1}]]\nno\\
v & \mapsto & Y^*_t(v,x) = \sum_{n \in {\Bbb
Z}}(v \otimes t^n)^*x^{-n-1}
 \end{eqnarray}
Then for general $Z(\cdot, x)$
as above, we can define a linear map
\begin{eqnarray}
\varepsilon_{Z}: \ V \otimes {\Bbb C}[t,t^{-1}]
& \rightarrow & U\nno\\
v \otimes t^n & \mapsto & v_{(n)}
\end{eqnarray}
(``evaluation with respect to $Z$''), i.e.,
\begin{equation}
\varepsilon_{Z}:\ Y_t(v,x) \mapsto Z(v,x),
\end{equation}
 and a linear map
\begin{eqnarray}
\varepsilon^*_{Z}:\ V \otimes {\Bbb C}[t,t^{-1}]
& \rightarrow & U\nno\\
v \otimes t^n & \mapsto & v^*_{(n)},
\end{eqnarray}
 i.e.,
\begin{equation}
\varepsilon^*_{Z}:\ Y_t(v,x) \mapsto
Z^*(v,x).
\end{equation}
 Then
\begin{equation}
\varepsilon^*_{Z} = \varepsilon_{Z} \circ *,
\end{equation}
that is,
\begin{equation}
\varepsilon_{Z}(Y^*_t(v,x)) = Z^*(v,x),
\end{equation}
or equivalently, the diagram
\begin{eqnarray}
Y_t(v,x) & \stackrel{\varepsilon_{Z}}{\longmapsto}
& Z(v,x)\nno\\
{*} \bar{\downarrow}\hspace{1.5em} & & \hspace{1.5em}
\bar{\downarrow} \;(Z(\cdot, x) \mapsto Z^*(\cdot, x))\nno\\
Y^*_t(v,x) &
\stackrel{\varepsilon_{Z}}{\longmapsto} & Z^*(v,x)
\end{eqnarray}
commutes. Note that the components $v^*_{(n)}$ of $Z^{*}(v, x)$ depend on
all the components $v_{(n)}$ of $Z(v,z)$ (for
arbitrary $v$), whereas the component $(v \otimes t^n)^*$ of
$Y^*_t(v,z)$ can be defined generically and abstractly; $(v
\otimes t^n)^*$ depends linearly on $v \in V$ alone.

     Since in general $Z^{**}(v, x) = Z(v, x)$, we know that
\begin{equation}
Y^{**}_t(v, x) =Y_t(v, x)
\end{equation}
as a special case, and in particular (and equivalently),
\begin{equation}
(v \otimes t^n)^{**} = v \otimes t^n
\end{equation}
for all $v\in V$ and
$n\in {\Bbb Z}$. Thus $*$ is an involution of $V \otimes {\Bbb
C}[t,t^{-1}]$.

     Furthermore, the involution $*$ of $V \otimes {\Bbb
C}[t,t^{-1}]$ extends canonically to a linear map $$V \otimes
{\Bbb C}[[t,t^{-1}]] \stackrel{*}{\rightarrow} V \otimes {\Bbb
C}[[t,t^{-1}]].$$ In fact, consider the restriction of $*$ to
$V=V \otimes t^0$:
\begin{eqnarray}
V &\stackrel{*}{\rightarrow} &V \otimes
{\Bbb C}[t,t^{-1}]\nno\\
v &\mapsto& v^* = (-1)^h \sum_{m \in {\Bbb N}}
\frac{1}{m!}(L(1)^m v) \otimes t^{-m-2+2h},
\end{eqnarray}
 extended by
linearity {from} $V_{(h)}$ to $V$.  Then for $v\in V$ and $n\in {\Bbb Z}$,
\begin{equation}
(v \otimes t^n)^* = v^*t^{-n},
\end{equation}
 and it is clear that $*$ extends to $V\otimes {\Bbb C}[[t, t^{-1}]]$:
For $f(t) \in {\Bbb C}[[t,t^{-1}]],$
\begin{equation}
(v \otimes
f(t))^* = v^*f(t^{-1}).
\end{equation}

To see that $*$ is an involution of this larger space, first note that
\begin{equation}
v^{**} = v
\end{equation}
(although $v^{*}\not\in V$ in general). (This could of course alternatively
be proved by direct calculation using formula (3.38).)
 Also, for $g(t) \in {\Bbb
C}[t,t^{-1}]$ and $f(t) \in {\Bbb C}[[t,t^{-1}]],$
\begin{equation}
(v \otimes
g(t)f(t))^* = v^*g(t^{-1})f(t^{-1})= (v \otimes
g(t))^*f(t^{-1}).
\end{equation}
 Thus for all $x \in V\otimes {\Bbb C}[t,t^{-
1}]$ and $f(t) \in {\Bbb C}[[t,t^{-1}]],$
\begin{equation}
(xf(t))^* = x^*f(t^{-
1}).
\end{equation}
  It follows that
\begin{eqnarray}
(v \otimes f(t))^{**} &
= & (v^*f(t^{-1}))^*\nno\\
& = & v^{**}f(t)\nno\\
& = & vf(t)\nno\\
& = & v
\otimes f(t),
\end{eqnarray}
and we have shown that $*$ is an involution of $V
\otimes {\Bbb C}[[t,t^{-1}]]$.
We have
\begin{equation}
*: V \otimes {\Bbb C}((t)) \leftrightarrow V
\otimes {\Bbb C}((t^{-1})).
\end{equation}

 Note that
\begin{eqnarray}
Y^*_t(v,x) & = & \sum_{n \in {\Bbb Z}}(v \otimes
t^n)^*x^{-n-1}\nno\\
& = & v^* \sum_{n \in {\Bbb Z}}t^{-n}x^{-n-1}\nno\\
&= & v^*x^{-1} \delta(tx)\nno\\
& = & v^*t \delta(tx)\nno\\
& \in & V
\otimes {\Bbb C}[[t,t^{-1},x,x^{-1}]].
\end{eqnarray}
Thus the map $v \mapsto Y^*_t(v,x)$ is the linear map given by
multiplying $v^*$ by the ``universal element'' $t \delta(tx)$
(cf. the comment following (3.6)).
For all $f(x) \in {\Bbb C}[[x,x^{-1}]],$ $f(x)Y^*_t(v,x)$
is defined and
\begin{eqnarray}
f(x)Y^*_t(v,x) & = & f(t^{-
1})Y^*_t(v,x)\nno\\
& = & v^*f(t^{-1})t\delta(tx).
\end{eqnarray}

     Now we return to the starting point --- the original special case:
 $U = \mbox{End}\; W$ and
$Z(\cdot,z) = Y_W(\cdot,z):\ V \to (\mbox{End}\; W)[[x,x^{-1}]].$
The corresponding map
\begin{eqnarray}
\varepsilon_{Z} =
\varepsilon_{Y_W}:\ V[t,t^{-1}] & \to & \mbox{End}\; W\nno\\
v \otimes
t^n & \mapsto & v_{(n)}
\end{eqnarray}
(recall (3.40)) is just the map
$\tau_W: v \otimes t^n
\mapsto v_n$ (recall (3.1)), i.e.,
 $v_{(n)} = v_n$ in this case.  Recall that this
map extends canonically to $V \otimes {\Bbb C}((t))$.  The map
$\varepsilon^*_{Z}$  is
$\tau_{W}\circ *:\ V \otimes {\Bbb C}[t,t^{-1}] \to \mbox{End}\;
W$ and this map extends canonically to $V \otimes {\Bbb C}((t^{-
1}))$. In addition to (3.7),  we have
\begin{eqnarray}
\tau_W(Y^*_t(v,z)) & = & Y^*_W(v,z)
\end{eqnarray}
($v^*_{(n)} =
v^*_n$ in this case; recall (3.21)).  In case $f(x) \in {\Bbb C}((x^{-1})),$
$$f(x)Y^*_W(v,x) = \tau_W(f(x)Y^*_t(v,x))$$
 is defined and is equal to
$\tau_W(f(t^{-1})Y^*_t(v,z))$ (which is also defined).

     The $x$-expansion coefficients of $f(x)Y^*_t(v,x)$, for
$f(x) \in {\Bbb C}[[x,x^{-1}]]$, span
\begin{eqnarray}
v^*f(t^{-
1}){\Bbb C}[t,t^{-1}] & = & (v{\Bbb C}[t,t^{-1}])^*f(t^{-1})\nno\\
& = &
(vf(t){\Bbb C}[t,t^{-1}])^*
\end{eqnarray}
The $x$-expansion coefficients of $Y^*_W(v,x)$ span
\begin{eqnarray}
\tau_W(v^*{\Bbb C}[t,t^{-1}])&=& \tau_W((v \otimes {\Bbb
C}[t,t^{-1}])^*)\nno\\
&=& \tau^*_W(v \otimes {\Bbb C}[t,t^{-
1}]).
\end{eqnarray}
In case $f(x) \in {\Bbb C}((x^{-1}))$, the $x$-expansion
coefficients of $f(x)Y^*_W(v,x)$ span $\tau_W(v^*f(t^{-1}){\Bbb
C}[t,t^{-1}])= \tau^*_W(vf(t){\Bbb C}[t,t^{-1}])$.
(Cf. the comments after (3.9).)

Our action of the space $V\otimes {\Bbb C}[t, t^{-1}, (z+t)^{-1}]$ will
be based on certain translation operations and on the $*$-operation.
More precisely, it is the space $V\otimes \iota_{+} {\Bbb C}[t, t^{-1},
(z+t)^{-1}]$ whose action we shall define, where we use the notations
\begin{eqnarray}
\iota_+:
{\Bbb C}(t) & \hookrightarrow & {\Bbb C}((t)) \subset {\Bbb
C}[[t,t^{-1}]]\nno\\
 \iota_-:{\Bbb C}(t) & \hookrightarrow & {\Bbb
C}((t^{-1})) \subset {\Bbb C} [[t,t^{-1}]]
\end{eqnarray}
to denote the operations of expanding a rational function of the
variable $t$ in
the indicated directions (as in Section 8.1 of
\cite{FLM2}).  For $a \in {\Bbb C}$, we define the translation
isomorphism
\begin{eqnarray}
T_a:\ {\Bbb C}(t) & \stackrel{\sim}{\rightarrow}
& {\Bbb C}(t)\nno\\f(t) & \mapsto & f(t+a)
\end{eqnarray}
and we set
\begin{equation}
T^\pm_a = \iota_\pm \circ T_a \circ
\iota^{-1}_+:\
\iota_+{\Bbb C}(t) \hookrightarrow {\Bbb C}((t^{\pm1})).
\end{equation}
(Note that the domains of these maps consist of certain series expansions
of rational functions rather than of rational functions themselves.)
 We shall be interested in
\begin{equation}
T^\pm_{-z}: \iota_+{\Bbb C}[t,t^{-1},(z +
t)^{-1}]
\hookrightarrow {\Bbb C}((t^{\pm1})),
\end{equation}
where $z$ is an arbitrary nonzero complex number.
The images of these two maps are $\iota_{\pm}{\Bbb C}[t,t^{-1},(z-t)^{-1}]$.

Extend $T^\pm_{-z}$ to linear isomorphisms
\begin{equation}
T^\pm_{-z}:\ V \otimes
\iota_+{\Bbb C}[t,t^{-1},(z+t)^{-1}] \stackrel{\sim}{\rightarrow}
V
\otimes \iota_\pm{\Bbb C}[t,t^{-1},(z-t)^{-1}]
\end{equation}
 given by $1\otimes T^\pm_{-z}$ with $T^\pm_{-z}$ as defined above.
Note that the domain of these two maps is described by (3.12)--(3.13),
that the image of the map $T^{+}_{-z}$ is described by (3.15)--(3.16)
and that the image of the map $T^{-}_{-z}$ is described by
(3.18)--(3.19).

We have the two mutually inverse maps
\begin{eqnarray}
V \otimes \iota_-{\Bbb C}[t,t^{-
1},(z-t)^{-1}] & \stackrel{*}{\rightarrow}&
V \otimes \iota_+{\Bbb C}[t,t^{-1},(z^{-1}-t)^{-1}]\nno\\
v \otimes f(t) &\mapsto &v^*f(t^{-1})
\end{eqnarray}
 and
\begin{eqnarray}
V \otimes
\iota_+{\Bbb
C}[t,t^{-1},(z^{-1}-t)^{-1}] & \stackrel{*}{\rightarrow}&
V \otimes \iota_-{\Bbb C}[t,t^{-1},(z-t)^{-1}]\nno\\
v \otimes
f(t) & \mapsto &v^*f(t^{-1}),
\end{eqnarray}
which are both isomorphisms. We form the composition
\begin{equation}
T^*_{-z} = *
\circ T^-_{-z}
\end{equation}
to obtain another isomorphism
$$T^*_{-z}: V \otimes \iota_+{\Bbb
C}[t,t^{-1},(z+t)^{-1}] \stackrel{\sim}{\rightarrow}
V
\otimes \iota_+{\Bbb C}[t,t^{-1},(z^{-1}-t)^{-1}].$$
The maps $T^{+}_{-z}$ and
$T^*_{-z}$ will be the main ingredients of our action. The following result
asserts that $T^{+}_{-z}$, $T^{-}_{-z}$ and
$T^*_{-z}$ transform the expression (3.12) into (3.15), (3.18) and
the $*$-transform of (3.18), respectively:
\begin{lemma} We have
\begin{eqnarray}
T_{-z}^{+}\left(z^{-1}\delta\left(\frac{x_{1}-x_{0}}{z}\right)Y_{t}(v,
x_{0})\right)
=x_{0}^{-1}\delta\left(\frac{z-x_{1}}{-x_{0}}\right) Y_{t}(v, x_{1}),\\
T_{-z}^{-}\left(z^{-1}\delta\left(\frac{x_{1}-x_{0}}{z}\right)Y_{t}(v,
x_{0})\right)
=x_{0}^{-1}\delta\left(\frac{x_{1}-z}{x_{0}}\right) Y_{t}(v, x_{1}),\\
T_{-z}^{*}\left(z^{-1}\delta\left(\frac{x_{1}-x_{0}}{z}\right)Y_{t}(v,
x_{0})\right)
=x_{0}^{-1}\delta\left(\frac{x_{1}-z}{x_{0}}\right) Y^{*}_{t}(v, x_{1}).
\end{eqnarray}
\end{lemma}
{\it Proof}\hspace{2ex} We  prove (3.71): {From} (3.12), the coefficient of
$x_{0}^{-n-1}x_{1}^{-m-1}$ in the left-hand side of
(3.71) is $T_{-z}^{+}(v\otimes (z+t)^{m}t^{n})$. By the definitions,
\begin{eqnarray}
T^{+}_{-z}(v\otimes (z+t)^{m}t^{n})
=v\otimes t^{m}(-(z-t))^{n}.
\end{eqnarray}
On the other hand, the right-hand side of (3.71) can be written as
\begin{eqnarray}
v\otimes
x_{0}^{-1}\delta\left(\frac{z-x_{1}}{-x_{0}}\right)  x_{1}^{-1}
\delta\left(\frac{t}{x_{1}}\right)
=v\otimes x_{0}^{-1}\delta\left(\frac{z-t}{-x_{0}}\right) x_{1}^{-1}
\delta\left(\frac{t}{x_{1}}\right)
\end{eqnarray}
where we have used (3.5) and the fundamental property (2.4) of the
formal $\delta$-function. The coefficient of
$x_{0}^{-n-1}x_{1}^{-m-1}$ in the right-hand side of (3.75) is also
$v\otimes t^{m}(-(z-t))^{n}$,  proving (3.71).  Formula
(3.72) is proved similarly, and (3.73) is obtained {from} (3.72) by the
application of the map $*$.
\epf

\renewcommand{\theequation}{\thesection.\arabic{equation}}
\renewcommand{\therema}{\thesection.\arabic{rema}}
\setcounter{equation}{0}
\setcounter{rema}{0}

\section{The notions of $P(z)$- and $Q(z)$-tensor product of two
modules}

For any ${\Bbb C}$-graded vector space $W=\coprod W_{(n)}$ such that
$\dim W_{(n)}<\infty$ for each $n\in {\Bbb C}$, we use the notation
\begin{equation}
\overline{W}=\prod_{n\in {\Bbb C}}W_{(n)}=W^{\prime *},
\end{equation}
where as usual $'$ denotes the graded dual space and $^{*}$ denotes
the dual space of a vector space.

Let $V$ be a vertex operator algebra and $W$ a $V$-module. For any
$v\in V$ and $n\in {\Bbb Z}$, $v_{n}$ acts naturally on $\overline{W}$
because of (2.21) for modules (recall Remark 2.8) and $v_{n}^{*}$ also
acts natually on $\overline{W}$, in view of (2.21) and (3.20).
Moreover, because of (2.21) and (2.44), for fixed $v\in V$, any
infinite linear combination of the $v_{n}$ of the form
$\sum_{n<N}a_{n}v_{n}$ ($a_{n}\in {\Bbb C}$) acts on $\overline{W}$,
and {from} (3.22), for example, we see that any infinite linear
combination of the form $\sum_{n>N}a_{n}v_{n}^*$ also acts on
$\overline{W}$.

Fix a nonzero complex number $z$ and let $(W_{1}, Y_{1})$ and $(W_{2},
Y_{2})$ be two $V$-modules. In the present paper (Part I), we give the
algebraic definitions and algebraic constructions of certain tensor
products of $(W_{1}, Y_{1})$ and $(W_{2}, Y_{2})$, depending on $z$,
but these have geometric meanings as well, which will be discussed and
studied in other papers in this series.  Let $(W_{3}, Y_{3})$ be
another $V$-module.  We call a {\it $P(z)$-intertwining map of type
${W_{3}}\choose {W_{1}W_{2}}$} (see Remark 4.3 below for the meaning
of the symbol $P(z)$) a linear map $F: W_{1}\otimes W_{2} \to
\overline{W}_{3}$ satisfying the condition
\bea
\lefteqn{x_{0}^{-1}\delta\left(\frac{ x_{1}-z}{x_{0}}\right)
Y_{3}(v, x_{1})F(w_{(1)}\otimes w_{(2)})=}\nonumber\\
&&=z^{-1}\delta\left(\frac{x_{1}-x_{0}}{z}\right)
F(Y_{1}(v, x_{0})w_{(1)}\otimes w_{(2)})\nonumber\\
&&\hspace{2em}+x_{0}^{-1}\delta\left(\frac{z-x_{1}}{-x_{0}}\right)
F(w_{(1)}\otimes Y_{2}(v, x_{1})w_{(2)})
\eea
for $v\in V$, $w_{(1)}\in W_{1}$, $w_{(2)}\in W_{2}$ (cf. the identity
(1.1) and the Jacobi identity (2.45) for intertwining operators). Note
that the left-hand side of (4.2) is well defined in view of the
comments in the preceding paragraph.  A {\it $P(z)$-product of $W_{1}$
and $W_{2}$} is a $V$-module $(W_{3}, Y_{3})$ equipped with a
$P(z)$-intertwining map $F$ of type ${W_{3}}\choose {W_{1}W_{2}}$.  We
denote it by $(W_{3}, Y_{3}; F)$ (or simply by $(W_{3}, F)$).  Let
$(W_{4}, Y_{4}; G)$ be another $P(z)$-product of $W_{1}$ and $W_{2}$.
A {\it morphism} {from} $(W_{3}, Y_{3}; F)$ to $(W_{4}, Y_{4}; G)$ is a
module map $\eta$ {from} $W_{3}$ to $W_{4}$ such that
\begin{equation}
G=\overline{\eta}\circ F,
\end{equation}
where $\overline{\eta}$ is the natural map {from} $\overline{W}_{3}$ to
$\overline{W}_{4}$ uniquely extending $\eta$.
 We define the notion
of $P(z)$-tensor product using a universal property as follows:
\begin{defi}
{\rm A {\it $P(z)$-tensor product of $W_{1}$ and $W_{2}$} is
a $P(z)$-product $(W_{1}\boxtimes_{P(z)} W_{2}, Y_{P(z)};
\boxtimes_{P(z)})$
such that for any $P(z)$-product
$(W_{3}, Y_{3}; F)$, there is a unique morphism {from}
$(W_{1}\boxtimes_{P(z)} W_{2}, Y_{P(z)};
\boxtimes_{P(z)})$ to $(W_{3}, Y_{3}; F)$.
The $V$-module $(W_{1}\boxtimes_{P(z)} W_{2},  Y_{P(z)})$ is
called a {\it $P(z)$-tensor product module} of $W_{1}$ and $W_{2}$.}
\end{defi}

\begin{rema}
{\rm As in the case of tensor products of modules for a Lie algebra, it
is clear {from} the definition that if a $P(z)$-tensor product of $W_{1}$
and $W_{2}$ exists, then it is unique up to unique isomorphism.}
\end{rema}

\begin{rema}
{\rm The symbol $P(z)$ in the definitions above in fact represents a geometric
object. Geometrically, to define a tensor product of $W_{1}$ and
$W_{2}$, we need to specify an element of the moduli space $K$ of
spheres with punctures and local coordinates vanishing at these
punctures. (In this remark and in Remark 4.6 below, we invoke the
detailed discussion of the moduli space $K$ and its role in the
geometric interpretation of the notion of vertex operator algebra
given in \cite{H1},
\cite{H2},
\cite{H3} or \cite{H4}. The present remark and Remark 4.6 are not
actually needed in the algebraic treatment in Part I.)  More
precisely, we need to specify an element of the determinant line
bundle over $K$ raised to the power $c$, where $c$ is the central
charge of the vertex operator algebra.  The definitions of
intertwining map, product and tensor product above are those
associated to the element $P(z)$ of $K$ containing ${\Bbb C}\cup \{
\infty \}$ with ordered punctures $\infty$, $z$, $0$ and standard
local coordinates $1/w$, $w-z$, $w$, vanishing at $\infty$, $z$, $0$,
respectively. Note that $P(z)$ is the geometric object corresponding
to vertex operators or intertwining operators in the geometric
interpretation of vertex operators and intertwining operators.  The
appropriate language describing tensor products defined using elements
of $K$ is that of operads, or more precisely, partial operads (see
\cite{M}, \cite{HL2},
\cite{HL3} and \cite{H5}).
These different tensor products will play important
roles in the formulations and constructions of the associativity and
commutativity isomorphisms.  }
\end{rema}

 Though it is natural to first consider $P(z)$-tensor products of two
modules as defined above, in this paper (Part I) we shall instead
construct another type of tensor product, the $Q(z)$-tensor product
(see below), since the calculations involved in the direct
construction of $Q(z)$-tensor products are simpler than those for
$P(z)$-tensor products.  Moreover, $P(z)$-tensor products can be
obtained {from} $Q(z)$-tensor products by performing certain geometric
transformations. We shall give the construction of a $P(z)$-tensor
product in Part III using the construction of the $Q(z)$-tensor product
presented in Sections 5 and 6 below. The reader should observe that many of
the considerations below concerning concepts based on $Q(z)$ carry over
immediately to the analogous considerations related to $P(z)$; in Part I
we focus only on $Q(z)$.

A {\it $Q(z)$-intertwining map of type ${W_{3}}\choose {W_{1}W_{2}}$}
is a linear map
$F: W_{1}\otimes W_{2} \to \overline{W}_{3}$ such that
\bea
\lefteqn{z^{-1}\delta\left(\frac{x_{1}-x_{0}}{z}\right)
Y^{*}_{3}(v, x_{0})F(w_{(1)}\otimes w_{(2)})=}\nonumber\\
&&=x_{0}^{-1}\delta\left(\frac{x_{1}-z}{x_{0}}\right)
F(Y_{1}^{*}(v, x_{1})w_{(1)}\otimes w_{(2)})\nonumber\\
&&\hspace{2em}-x_{0}^{-1}\delta\left(\frac{z-x_{1}}{-x_{0}}\right)
F(w_{(1)}\otimes Y_{2}(v, x_{1})w_{(2)})
\eea
for $v\in V$, $w_{(1)}\in W_{1}$, $w_{(2)}\in W_{2}$. As in the
definition of $P(z)$-intertwining map, note that the left-hand side
and both terms on the right-hand side of (4.4) are well defined.
First replacing $v$ by $(-x_{0}^{2})^{L(0)}e^{-x_{0}L(1)}v$ and then
replacing $x_{0}$ by $x_{0}^{-1}$ in (4.4), we see that (4.4) is
equivalent to:
\bea
\lefteqn{z^{-1}\delta\left(\frac{x_{1}-x^{-1}_{0}}{z}\right)
Y_{3}(v, x_{0})F(w_{(1)}\otimes w_{(2)})=}\nonumber\\
&&=x_{0}\delta\left(\frac{x_{1}-z}{x^{-1}_{0}}\right)
F(Y_{1}(e^{x_{1}L(1)}(x_{1}x_{0})^{-2L(0)}e^{-x^{-1}_{0}L(1)}v, x^{-1}_{1})
w_{(1)}\otimes w_{(2)})\nonumber\\
&&\hspace{2em}-x_{0}\delta\left(\frac{z-x_{1}}{-x^{-1}_{0}}\right)
F(w_{(1)}\otimes Y_{2}((-x_{0}^{-2})^{L(0)}e^{-x^{-1}_{0}L(1)}v,
x_{1})w_{(2)}).
\eea
(The reverse procedure is given by first inverting $x_{0}$ and then
replacing $v$ by $e^{x_{0}L(1)}(-x_{0}^{-2})^{L(0)}v$.)

We denote the vector space of $Q(z)$-intertwining maps of type
${W_{3}}\choose {W_{1}W_{2}}$ by ${\cal M}[Q(z)]^{W_{3}}_{W_{1}W_{2}}$ or
simply by ${\cal M}^{W_{3}}_{W_{1}W_{2}}$.

We define a {\it $Q(z)$-product of $W_{1}$ and $W_{2}$} to be a
$V$-module $(W_{3}, Y_{3})$ together with a $Q(z)$-intertwining map
$F$ of type ${W_{3}}\choose {W_{1}W_{2}}$ and we denote it by $(W_{3},
Y_{3}; F)$ (or $(W_{3}, F)$).
 Let $(W_{3}, Y_{3}; F)$ and $(W_{4}, Y_{4}; G)$ be two
$Q(z)$-products of $W_{1}$ and $W_{2}$. A {\it morphism} {from}
$(W_{3}, Y_{3}; F)$ to $(W_{4}, Y_{4}; G)$ is a module map $\eta$ {from}
$W_{3}$ to $W_{4}$ such that
\begin{equation}
G=\overline{\eta}\circ F
\end{equation}
where, as in (4.3), $\overline{\eta}$ is the natural map {from}
$\overline{W}_{3}$ to $\overline{W}_{4}$ uniquely extending $\eta$.

\begin{defi}
{\rm A {\it $Q(z)$-tensor product of $W_{1}$ and $W_{2}$} is
a $Q(z)$-product $(W_{1}\boxtimes_{Q(z)} W_{2},
Y_{Q(z)}; \boxtimes_{Q(z)})$
such that for any $Q(z)$-product
$(W_{3}, Y_{3}; F)$, there is a unique morphism {from}
$(W_{1}\boxtimes_{Q(z)} W_{2}, Y_{Q(z)};
\boxtimes_{Q(z)})$ to $(W_{3}, Y_{3}; F)$.
The $V$-module $(W_{1}\boxtimes_{Q(z)} W_{2},  Y_{Q(z)})$ is
called a {\it $Q(z)$-tensor product module} of $W_{1}$ and $W_{2}$.}
\end{defi}

\begin{rema}
{\rm As in the case of $P(z)$-tensor products, a $Q(z)$-tensor product
is unique up to unique isomorphism if it exists. }
\end{rema}

\begin{rema}
{\rm In the definitions above, $Q(z)$ represents the element of $K$
containing ${\Bbb C}\cup \{ \infty \}$ with ordered punctures $z$,
$\infty$, $0$ and standard local coordinates vanishing at these
punctures. (Recall Remark 4.3.) In fact, this is the same as the
element of $K$ containing ${\Bbb C}\cup \{ \infty \}$ with ordered
punctures $\infty$, $1/z$, $0$ and local coordinates $z/(zw-1)$,
$(zw-1)/z^{2}w$, $z^{2}w/(zw-1)$ vanishing at $\infty$, $1/z$, $0$,
respectively, and (4.5) corresponds to this canonical sphere with
punctures and local coordinates. }
\end{rema}

The existence of
a $Q(z)$-tensor product  is not obvious. We shall prove the existence and
give two
constructions under
certain assumptions on the vertex operator algebra in this and
the next two sections.
First we relate $Q(z)$-intertwining maps of type ${W_{3}}\choose {W_{1}W_{2}}$
to intertwining operators of type ${W'_{1}}\choose {W'_{3}W_{2}}$.

Let ${\cal Y}$ be an intertwining operator of type ${W_{3}}\choose
{W_{1}W_{2}}$. For any complex number $\zeta$ and any $w_{(1)}\in
W_{1}$, ${\cal Y}(w_{(1)}, x)\lbar_{x^{n}=e^{n\zeta},\ n\in {\Bbb C}}$
is a well-defined map {from} $W_{2}$ to $\overline{W}_{3}$, in view of
formula (2.21) for intertwining operators. For brevity
of notation, we shall write this map as ${\cal Y}(w_{(1)},
e^{\zeta})$, but note that ${\cal Y}(w_{(1)}, e^{\zeta})$ depends on
$\zeta$, not on just $e^{\zeta}$, as the notation might suggest.  In
this paper we shall always choose $\log z$ so that
\begin{equation}
\log z=\log |z|+i\arg z\;\;\mbox{\rm with}\;\;0\le\arg z<2\pi.
\end{equation}
Arbitrary values of the $\log$ function will be denoted
\begin{equation}
l_{p}(z)=\log z+2p\pi i
\end{equation}
for $p\in {\Bbb Z}$.

We now describe the close connection between
 intertwining operators of type ${W'_{1}}\choose {W'_{3}W_{2}}$
and $Q(z)$-intertwining maps of type ${W_{3}}\choose {W_{1}W_{2}}$.
Fix an integer $p$. Let ${\cal Y}$ be an intertwining operator of
type ${W'_{1}}\choose {W'_{3}W_{2}}$. Then we have an element of
$(W_{1}\otimes W'_{3}\otimes W_{2})^{*}$ whose value at
$w_{(1)}\otimes w'_{(3)}\otimes w_{(2)}$ ($w_{(1)}\in W_{1}$,
$w_{(2)}\in W_{2}$, $w'_{(3)}\in W'_{3}$) is
$$\langle w_{(1)},
{\cal Y}(w'_{(3)}, e^{l_{p}(z)})w_{(2)}\rangle_{W'_{1}}$$
where $\langle \cdot, \cdot \rangle_{W'_{1}}$ is the pairing between
$W_{1}$ and $\overline{W'_{1}}=W^{*}_{1}$.
Since any element of $(W_{1}\otimes W'_{3}\otimes W_{2})^{*}$ amounts
exactly to a
linear map {from} $W_{1}\otimes W_{2}$ to $W_{3}^{\prime *}=\overline{W}_{3}$,
our element of $(W_{1}\otimes W'_{3}\otimes W_{2})^{*}$ obtained {from} the
intertwining operator ${\cal Y}$ gives us a linear map
$F_{{\cal Y}, p}: W_{1}\otimes W_{2}\to \overline{W}_{3}$ such that
\begin{equation}
\langle w'_{(3)}, F_{{\cal Y}, p}(w_{(1)}\otimes w_{(2)})\rangle_{W_{3}}
=\langle w_{(1)}, {\cal Y}(w'_{(3)}, e^{l_{p}(z)})w_{(2)}\rangle_{W'_{1}}
\end{equation}
for all $w_{(1)}\in W_{1}$, $w_{(2)}\in W_{2}$, $w'_{(3)}\in W'_{3}$, where
$\langle \cdot, \cdot \rangle_{W_{3}}$ is the pairing between $W'_{3}$ and
$\overline{W}_{3}$. (For any module $W$, we shall use the analogous
notation $\langle \cdot, \cdot \rangle_{W}$ to denote the pairing
between $W'$ and $\overline{W}$.)
The Jacobi identity for ${\cal Y}$ is equivalent to the identity
\begin{eqnarray}
\lefteqn{x_{2}^{-1}\delta\left(\frac{x_{1}-x_{0}}{x_{2}}\right)
\langle w_{(1)}, {\cal Y}(Y'_{3}(v, x_{0})
w'_{(3)}, x_{2})w_{(2)}\rangle_{W'_{1}}}\nonumber\\
&&=x^{-1}_{0}\delta\left(\frac{x_{1}-x_{2}}{x_{0}}\right)
\langle
w_{(1)},  Y'_{1}(v, x_{1}){\cal Y}(w'_{(3)}, x_{2})w_{(2)}
\rangle_{W'_{1}}\nonumber\\
&&\hspace{2em}-x_{0}^{-1}\delta\left(\frac{x_{2}-x_{1}}{-x_{0}}\right)
\langle w_{(1)}, {\cal Y}(w'_{(3)}, x_{2})
Y_{2}(v, x_{1})
w_{(2)}\rangle_{W'_{1}}
\end{eqnarray}
for all $w_{(1)}$, $w_{(2)}$ and $w'_{(3)}$ (recall the notation (2.48)).
Substituting  $e^{nl_{p}(z)}$ for $x^{n}_{2}$, $n\in {\Bbb C}$, in (4.10), and
noting that in case $n\in {\Bbb Z}$, we may simply write $z^{n}$ for
$e^{nl_{p}(z)}$,
we obtain
\begin{eqnarray}
\lefteqn{z^{-1}\delta\left(\frac{x_{1}-x_{0}}{z}\right)
\langle w_{(1)}, {\cal Y}(Y'_{3}(v, x_{0})
w'_{(3)}, e^{l_{p}(z)})w_{(2)}\rangle_{W'_1}}\nonumber\\
&&=x^{-1}_{0}\delta\left(\frac{x_{1}-z}{x_{0}}\right)\langle w_{(1)},
Y'_{1}(v, x_{1}){\cal Y}(w'_{(3)}, e^{l_{p}(z)})w_{(2)}
\rangle_{W'_1}\nonumber\\
&&\hspace{2em}-x_{0}^{-1}\delta\left(\frac{z-x_{1}}{-x_{0}}\right)
\langle w_{(1)}, {\cal Y}(w'_{(3)}, e^{l_{p}(z)})
Y_{2}(v, x_{1})w_{(2)}\rangle_{W'_1}.
\end{eqnarray}
Using (3.22) and (4.9), we see that (4.11) can be written as
\begin{eqnarray}
\lefteqn{z^{-1}\delta\left(\frac{x_{1}-x_{0}}{z}\right)
\langle
w'_{(3)}, Y^{*}_{3}(v, x_{0})F_{{\cal Y}, p}(w_{(1)}\otimes w_{(2)})
\rangle_{W_3}}
\nonumber\\
&&=x^{-1}_{0}\delta\left(\frac{x_{1}-z}{x_{0}}\right)\langle
w'_{(3)}, F_{{\cal Y}, p}(Y^{*}_{1}(v, x_{1})w_{(1)}\otimes w_{(2)})
\rangle_{W_3}
\nonumber\\
&&\hspace{2em}-x_{0}^{-1}\delta\left(\frac{z-x_{1}}{-x_{0}}\right)
\langle w'_{(3)}, F_{{\cal Y}, p}(w_{(1)}\otimes Y_{2}(v, x_{1})w_{(2)})
\rangle_{W_3}.
\end{eqnarray}
 Thus  $F_{{\cal Y}, p}$ is a $Q(z)$-intertwining map of
type ${W_{3}}\choose {W_{1}W_{2}}$.

The only part of the definition of intertwining operator we have not yet used
is the $L(-1)$-derivative property (2.46). (Recall that the lower truncation
condition (2.44) has already been used in the formulation of the first term
on the left-hand side of the Jacobi identity (2.45).) Since we have specialized
 $x$ to $z$ in ${\cal Y}(\cdot, x)$, there is no property of $F_{{\cal Y}, p}$
corresponding the $L(-1)$-derivative property of ${\cal Y}$. Instead, the
$L(-1)$-derivative property will enable us to recover ${\cal Y}(\cdot, x)$
{from} $F_{{\cal Y}, p}$. Specifically, the $L(-1)$-derivative property enters
into the proof of the formula
\begin{equation}
x^{L(0)}{\cal Y}(w'_{(3)}, x_{0})x^{-L(0)}={\cal Y}(x^{L(0)}w'_{(3)}, xx_{0})
\end{equation}
(recall \cite{FHL}, formula (5.4.22), and Lemma 5.2.3 and its proof),
and this is equivalent to the formula
\begin{equation}
\langle x^{L(0)}w_{(1)}, {\cal Y}(x^{-L(0)}w'_{(3)},
x_{0})x^{-L(0)}w_{(2)}\rangle_{W'_1}
=\langle w_{(1)}, {\cal Y}(w'_{(3)}, xx_{0})w_{(2)}\rangle_{W'_1}
\end{equation}
for all $w_{(1)}\in W_{1}$.
Substituting   $e^{nl_{p}(z)}$ for $x^{n}_{0}$ and
$e^{-nl_{p}(z)}x^{n}$ for $x^{n}$, $n\in {\Bbb C}$,
 in (4.14),
we obtain
\begin{eqnarray}
&\langle e^{-l_{p}(z)L(0)}x^{L(0)}w_{(1)},
{\cal Y}(e^{l_{p}(z)L(0)}x^{-L(0)}w'_{(3)}, e^{l_{p}(z)})
e^{l_{p}(z)L(0)}x^{-L(0)}w_{(2)}\rangle_{W'_1}&\nno\\
&=\langle w_{(1)}, {\cal Y}(w'_{(3)}, x)w_{(2)}\rangle_{W'_1},&
\end{eqnarray}
or equivalently, by (4.9),
\begin{eqnarray}
&\langle e^{l_{p}(z)L(0)}x^{-L(0)}w'_{(3)}, F_{{\cal Y}, p}
(e^{-l_{p}(z)L(0)}x^{L(0)}w_{(1)}
\otimes
e^{l_{p}(z)L(0)}x^{-L(0)}w_{(2)})\rangle_{W_3}&\nno\\
&=\langle w_{(1)}, {\cal Y}(w'_{(3)}, x)w_{(2)}\rangle_{W'_1}.&
\end{eqnarray}
Thus we have  recovered ${\cal Y}$ {from} $F_{{\cal Y}, p}$.

We shall also need the following alternative way of recovering ${\cal
Y}$ {from} $F_{{\cal Y}, p}$, using components. We write
(4.9) as:
\begin{equation}
\langle w_{(1)},
\sum_{n\in {\Bbb C}}(w'_{(3)})_{n}w_{(2)}e^{(-n-1)l_{p}(z)}\rangle_{W'_1}
=\langle w'_{(3)},
F_{{\cal Y}, p}(w_{(1)}\otimes w_{(2)})\rangle_{W_3}.
\end{equation}
This formula enables us to recover the components
$(w'_{(3)})_{n}w_{(2)}$, $n\in {\Bbb C}$, of ${\cal Y}(w'_{(3)},
x)w_{(2)}$ {from} $F_{{\cal Y}, p}$, assuming for convenience that
$w_{(2)}$ and $w'_{(3)}$ are homogeneous vectors, in the following
way: The map $F_{{\cal Y}, p}$ gives an element of $(W_{1}\otimes
W'_{3}\otimes W_{2})^{*}$ whose value at $w_{(1)}\otimes
w'_{(3)}\otimes w_{(2)}$ is equal to the right-hand side of (4.17).
This element amounts to a map {from} $W'_{3}\otimes W_{2}$ to
$W^{*}_{1}$. By (4.17), the image of $w'_{(3)}\otimes w_{(2)}$ under
this map is equal to $\sum_{n\in {\Bbb
C}}(w'_{(3)})_{n}w_{(2)}e^{(-n-1)l_{p}(z)}$. Projecting this image to
the homogeneous subspace of $W'_{1}$ of weight equal to $$\mbox{\rm
wt}\;w'_{(3)}-n-1+\mbox{\rm wt}\; w_{(2)},$$ we obtain
$(w'_{(3)})_{n}w_{(2)}e^{(-n-1)l_{p}(z)}$. Mutiplying this by
$e^{(n+1)l_{p}(z)}$, we recover the coefficient $(w'_{(3)})_{n}w_{(2)}$.

Motivated by this procedure, we would like to construct an
intertwining operator of type ${W'_{1}}\choose {W'_{3}W_{2}}$ {from} a
$Q(z)$-intertwining map of type ${W_{3}}\choose {W_{1}W_{2}}$. Let $F$
be a $Q(z)$-intertwining map of type ${W_{3}}\choose {W_{1}W_{2}}$.
This linear map {from} $W_{1}\otimes W_{2}$ to $\overline{W}_{3}$
gives us an element of $(W_{1}\otimes W'_{3}\otimes W_{2})^{*}$ whose
value at $w_{(1)}\otimes w'_{(3)}\otimes w_{(2)}$ is $$\langle
w'_{(3)}, F(w_{(1)}\otimes w_{(2)})\rangle_{W_3}.$$ But since every
element of $(W_{1}\otimes W'_{3}\otimes W_{2})^{*}$ also amounts to a
linear map {from} $W'_{3}\otimes W_{2}$ to $W^{*}_{1}$, we have such a
map as well. Let $w'_{(3)}\in W'_{3}$ and $w_{(2)}\in W_{2}$ be
homogeneous elements. Since $W^{*}_{1}=\prod_{n\in {\Bbb
C}}(W'_{1})_{(n)}$, the image of $w'_{(3)}\otimes w_{(2)}$ under our
map can be written as $\sum_{n\in {\Bbb
C}}(w'_{(3)})_{n}w_{(2)}e^{(-n-1)l_{p}(z)}$ where for any $n\in {\Bbb
C}$, $(w'_{(3)})_{n}w_{(2)}e^{(-n-1)l_{p}(z)}$ is the projection of
the image to the homogeneous subspace of $W'_{1}$ of weight equal to
$$\mbox{\rm wt}\;w'_{(3)}-n-1+\mbox{\rm wt}\; w_{(2)}.$$ (Here we are
defining elements denoted $(w'_{(3)})_{n}w_{(2)}$ of $W'_{1}$ for
$n\in {\Bbb C}$.)  We define
$${\cal Y}_{F,p}(w'_{(3)}, x)w_{(2)}
=\sum_{n\in {\Bbb C}}(w'_{(3)})_{n}w_{(2)}x^{-n-1}\in W'_{1}\{ x\}$$
for all homogeneous elements $w'_{(3)}\in W'_{3}$ and $w_{(2)}\in
W_{2}$. Using linearity, we extend ${\cal Y}_{F,p}$ to a linear map
\begin{eqnarray}
W'_{3}\otimes W_{2}&\to& W'_{1}\{ x\}\nno\\
w'_{(3)}\otimes w_{(2)}&\mapsto &{\cal Y}_{F,p}(w'_{(3)}, x)w_{(2)}.
\end{eqnarray}
The correspondence $F\mapsto {\cal Y}_{F, p}$ is linear, and {from}
the definitions and the discussion in the preceding paragraph,
we have ${\cal Y}_{F_{{\cal Y}, p}, p}={\cal Y}$ for an intertwining operator
${\cal Y}$ of
type ${W'_{1}}\choose {W'_{3}W_{2}}$.

\begin{propo}
For $p\in {\Bbb Z}$, the correspondence ${\cal Y}\mapsto F_{{\cal Y}, p}$ is
a linear isomorphism {from} the space ${\cal
V}^{W'_{1}}_{W'_{3}W_{2}}$ of intertwining operators of type
${W'_{1}}\choose {W'_{3}\; W_{2}}$ to the space ${\cal
M}^{W_{3}}_{W_{1}W_{2}} ={\cal M}[Q(z)]^{W_{3}}_{W_{1}W_{2}}$ of
$Q(z)$-intertwining maps of type ${W_{3}}\choose {W_{1}W_{2}}$. Its
inverse is given by $F\mapsto {\cal Y}_{F, p}$.
\end{propo}
\pf
We need only show that for any
 $Q(z)$-intertwining map $F$ of type ${W_{3}}\choose {W_{1}W_{2}}$,
${\cal Y}_{F, p}$ is an intertwining operator of type
${W'_{1}}\choose {W'_{3}\; W_{2}}$. {From} the discussion above and
the definition of ${\cal Y}_{F, p}$, the lower truncation condition
(2.44) holds
for ${\cal Y}_{F, p}$ and
we have the  equality
\begin{eqnarray}
\lefteqn{\langle w_{(1)}, {\cal Y}_{F, p}(w'_{(3)}, x)
w_{(2)}\rangle_{W'_1}=}\nno\\
&&=\langle e^{l_{p}(z)L(0)}x^{-L(0)}w'_{(3)}, F
(e^{-l_{p}(z)L(0)}x^{L(0)}w_{(1)}\otimes\nno\\
&&\hspace{2em}\otimes e^{l_{p}(z)L(0)}x^{-L(0)}w_{(2)})\rangle_{W_3}
\end{eqnarray}
(cf. (4.16)). Now (4.4) gives
\bea
\lefteqn{z^{-1}\delta\left(\frac{x_{1}-x_{0}}{z}\right)
\langle w'_{(3)}, Y^{*}_{3}(v, x_{0})F(w_{(1)}\otimes
w_{(2)})\rangle_{W_3}=}
\nonumber\\
&&=x^{-1}_{0}\delta\left(\frac{x_{1}-z}{x_{0}}\right)
\langle w'_{(3)}, F(Y_{1}^{*}(v, x_{1})w_{(1)}\otimes w_{(2)})
\rangle_{W_3}\nno\\
&&\hspace{2em}-x_{0}^{-1}\delta\left(\frac{z-x_{1}}{-x_{0}}\right)
\langle w'_{(3)},F(w_{(1)}\otimes Y_{2}(v, x_{1})w_{(2)})\rangle_{W_3}.
\eea
Changing the formal variables $x_{0}$ and $x_{1}$ in (4.20) to
$zx_{2}^{-1}x_{0}$
and $zx_{2}^{-1}x_{1}$, respectively, and using (3.22), we obtain
\bea
\lefteqn{x_{2}^{-1}\delta\left(\frac{x_{1}-x_{0}}{x_{2}}\right)
\langle Y'_{3}(v, zx_{2}^{-1}x_{0})w'_{(3)}, F(w_{(1)}\otimes
w_{(2)})\rangle_{W_3}=}
\nonumber\\
&&=x^{-1}_{0}\delta\left(\frac{x_{1}-x_{2}}{x_{0}}\right)
\langle w'_{(3)}, F(Y_{1}^{*}(v, zx_{2}^{-1}x_{1})w_{(1)}\otimes w_{(2)})
\rangle_{W_3}\nno\\
&&\hspace{2em}-x_{0}^{-1}\delta\left(\frac{x_{2}-x_{1}}{-x_{0}}\right)
\langle w'_{(3)},F(w_{(1)}\otimes Y_{2}(v, zx_{2}^{-1}x_{1})w_{(2)})
\rangle_{W_3}.
\eea
(Note that all powers of $z$ occurring here are integral.)
Using the formulas
\begin{eqnarray}
Y'_{3}(v, zx_{2}^{-1}x_{0})&=&e^{l_{p}(z)L(0)}x_{2}^{-L(0)}
Y'_{3}(e^{-l_{p}(z)L(0)}x_{2}^{L(0)}v, x_{0})\cdot\nno\\
&&\hspace{6em}\cdot e^{-l_{p}(z)L(0)}
x_{2}^{L(0)},\\
Y_{1}^{*}(v, zx_{2}^{-1}x_{1})&=&e^{-l_{p}(z)L(0)}x_{2}^{L(0)}
Y_{1}^{*}(e^{-l_{p}(z)L(0)}x_{2}^{L(0)}v, x_{1})\cdot \nno\\
&&\hspace{6em}\cdot e^{l_{p}(z)L(0)}x_{2}^{-L(0)},\\
Y_{2}(v, zx_{2}^{-1}x_{1})&=&e^{l_{p}(z)L(0)}x_{2}^{-L(0)}
Y_{2}(e^{-l_{p}(z)L(0)}x_{2}^{L(0)}v, x_{1})\cdot \nno\\
&&\hspace{6em}\cdot e^{-l_{p}(z)L(0)}x_{2}^{L(0)},
\end{eqnarray}
which follow {from} Lemma 5.2.3 together with formula (5.2.39) of \cite{FHL}
(note that the eigenvalues of $L(0)$ are not in general integral on the
modules),
we see that (4.21) becomes
\bea
&&x_{2}^{-1}\delta\left(\frac{x_{1}-x_{0}}{x_{2}}\right)
\langle e^{l_{p}(z)L(0)}x_{2}^{-L(0)}
Y'_{3}(e^{-l_{p}(z)L(0)}x_{2}^{L(0)}v, x_{0})\cdot \nno\\
&&\hspace{8em}\cdot e^{-l_{p}(z)L(0)}x_{2}^{L(0)}w'_{(3)}, F(w_{(1)}\otimes
w_{(2)})\rangle_{W_3}
\nonumber\\
&&=x^{-1}_{0}\delta\left(\frac{x_{1}-x_{2}}{x_{0}}\right)
\langle w'_{(3)}, F(e^{-l_{p}(z)L(0)}x_{2}^{L(0)}
\cdot \nno\\
&&\hspace{3em}\cdot Y_{1}^{*}(e^{-l_{p}(z)L(0)}x_{2}^{L(0)}v, x_{1})
e^{l_{p}(z)L(0)}x_{2}^{-L(0)}
w_{(1)}\otimes w_{(2)})\rangle_{W_3}\nno\\
&&\hspace{2em}-x_{0}^{-1}\delta\left(\frac{x_{2}-x_{1}}{-x_{0}}\right)
\langle w'_{(3)},F(w_{(1)}\otimes e^{l_{p}(z)L(0)}x_{2}^{-L(0)}\cdot \nno\\
&&\hspace{3em}\cdot
Y_{2}(e^{-l_{p}(z)L(0)}x_{2}^{L(0)}v, x_{1})e^{-l_{p}(z)L(0)}x_{2}^{L(0)}
w_{(2)})\rangle_{W_3}.
\eea
Replacing $v$, $w_{(1)}$, $w_{(2)}$ and $w'_{(3)}$ in (4.25) by
$$e^{l_{p}(z)L(0)}x_{2}^{-L(0)}v,$$
$$e^{-l_{p}(z)L(0)}x_{2}^{L(0)}w_{(1)},$$
$$e^{l_{p}(z)L(0)}x_{2}^{-L(0)}w_{(2)}$$ and
$$e^{l_{p}(z)L(0)}x_{2}^{-L(0)}w'_{(3)},$$ respectively, we obtain
\bea
&&x_{2}^{-1}\delta\left(\frac{x_{1}-x_{0}}{x_{2}}\right)
\langle e^{l_{p}(z)L(0)}x_{2}^{-L(0)}
Y'_{3}(v, x_{0})
w'_{(3)}, \nno\\
&&\hspace{8em}F(e^{-l_{p}(z)L(0)}x_{2}^{L(0)}w_{(1)}
\otimes e^{l_{p}(z)L(0)}x_{2}^{-L(0)}w_{(2)})\rangle_{W_3}
\nonumber\\
&&=x^{-1}_{0}\delta\left(\frac{x_{1}-x_{2}}{x_{0}}\right)
\langle e^{l_{p}(z)L(0)}x_{2}^{-L(0)}w'_{(3)},
F(e^{-l_{p}(z)L(0)}x_{2}^{L(0)}\cdot \nno\\
&&\hspace{8em}\cdot Y_{1}^{*}(v, x_{1})
w_{(1)}\otimes e^{l_{p}(z)L(0)}x_{2}^{-L(0)}w_{(2)})\rangle_{W_3}\nno\\
&&\hspace{2em}-x_{0}^{-1}\delta\left(\frac{x_{2}-x_{1}}{-x_{0}}\right)
\langle e^{l_{p}(z)L(0)}x_{2}^{-L(0)}w'_{(3)},
F(e^{-l_{p}(z)L(0)}\cdot \nno\\
&&\hspace{8em}\cdot x_{2}^{L(0)}w_{(1)}
\otimes e^{l_{p}(z)L(0)}x_{2}^{-L(0)}
Y_{2}(v, x_{1})
w_{(2)})\rangle_{W_3}.
\eea
But using (4.19), we can write (4.26) as
\bea
\lefteqn{x_{2}^{-1}\delta\left(\frac{x_{1}-x_{0}}{x_{2}}\right)
\langle w_{(1)}, {\cal Y}_{F, p}(
Y'_{3}(v, x_{0})
w'_{(3)}, x_{2})w_{(2)}\rangle_{W'_1}}\nonumber\\
&&=x^{-1}_{0}\delta\left(\frac{x_{1}-x_{2}}{x_{0}}\right)
\langle Y_{1}^{*}(v, x_{1})
w_{(1)}, {\cal Y}_{F, p}(w'_{(3)}, x_{2})w_{(2)}\rangle_{W'_1}\nonumber\\
&&\hspace{2em}-x_{0}^{-1}\delta\left(\frac{x_{2}-x_{1}}{-x_{0}}\right)
\langle w_{(1)}, {\cal Y}_{F, p}(w'_{(3)}, x_{2})
Y_{2}(v, x_{1})
w_{(2)}\rangle_{W'_1},
\eea
and (4.27) is equivalent to the Jacobi identity
for ${\cal Y}_{F, p}$.

Finally, the Jacobi identity implies that
$$[L(0), {\cal Y}_{F, p}(w'_{(3)}, x)]={\cal Y}_{F, p}(L(0)w'_{(3)}, x)+
x{\cal Y}_{F, p}(L(-1)w'_{(3)}, x),$$
and since by construction the weight of the operator $(w'_{(3)})_{n}$
($n\in {\Bbb C}$) is wt$\ w'_{(3)}-n-1$ if $w'_{(3)}$ is homogeneous, the
$L(-1)$-derivative property follows.\epfv

The following immediate result relates module maps {from} a tensor
product module with intertwining maps and intertwining operators:
\begin{propo}
Suppose that $W_{1}\boxtimes_{Q(z)}W_{2}$ exists. We have a natural
isomorphism
\begin{eqnarray}
\mbox{\rm Hom}_{V}(W_{1}\boxtimes_{Q(z)}W_{2}, W_{3})&\stackrel{\sim}{\to}&
{\cal M}^{W_{3}}_{W_{1}W_{2}}\nno\\
\eta&\mapsto& \overline{\eta}\circ \boxtimes_{Q(z)}
\end{eqnarray}
and for $p\in {\Bbb Z}$, a natural isomorphism
\begin{eqnarray}
\mbox{\rm Hom}_{V}(W_{1}\boxtimes_{Q(z)} W_{2}, W_{3})&
\stackrel{\sim}{\rightarrow}&
{\cal V}^{W'_{1}}_{W'_{3}W_{2}}\nno\\
\eta&\mapsto & {\cal Y}_{\eta, p}
\end{eqnarray}
where ${\cal Y}_{\eta, p}={\cal Y}_{F, p}$ with
$F=\overline{\eta}\circ \boxtimes_{Q(z)}$.\epf
\end{propo}

In Part II we shall prove the following:

\begin{propo}
For any integer $r$, there is a natural isomorphism
\begin{equation}
B_{r}: {\cal V}^{W_{3}}_{W_{1}W_{2}}\to
{\cal V}^{W'_{1}}_{W'_{3}W_{2}}
\end{equation}
defined by the condition  that for any
intertwining operator ${\cal Y}$ in ${\cal V}^{W_{3}}_{W_{1}W_{2}}$ and
$w_{(1)}\in W_{1}$, $w_{(2)}\in W_{2}$, $w'_{(3)}\in W'_{3}$,
\begin{eqnarray}
\lefteqn{\langle w_{(1)}, B_{r}({\cal Y})(w'_{(3)}, x)
w_{(2)}\rangle_{W'_1}=}\nno\\
&&=\langle e^{-x^{-1}L(1)}w'_{(3)}, {\cal Y}(e^{xL(1)}w_{(1)},
x^{-1})\cdot\nno\\
&&\hspace{4em}\cdot e^{-xL(1)}e^{(2r+1)\pi iL(0)}
x^{-2L(0)}w_{(2)}\rangle_{W_3}.
\end{eqnarray}
\end{propo}

 Combining the last two results, we obtain:

\begin{corol}
 For any $V$-modules $W_{1}$, $W_{2}$, $W_{3}$ such that
$W_{1}\boxtimes_{Q(z)}W_{2}$ exists and any integers $p$ and $r$,
we have a natural isomorphism
\begin{eqnarray}
\mbox{\rm Hom}_{V}(W_{1}\boxtimes_{Q(z)} W_{2}, W_{3})&
\stackrel{\sim}{\rightarrow}&
{\cal V}^{W_{3}}_{W_{1}W_{2}}\nno\\
\eta&\mapsto &B^{-1}_{r}({\cal Y}_{\eta, p}).\hspace{2em}\Box
\end{eqnarray}
\end{corol}

It is clear {from} Definition 4.4 that the tensor product operation
distributes over direct sums in the following sense:

\begin{propo}
For $V$-modules $U_{1}, \dots, U_{k}$, $W_{1}, \dots, W_{l}$, suppose
that each $U_{i}\boxtimes_{Q(z)}W_{j}$ exists. Then
$(\coprod_{i}U_{i})\boxtimes_{Q(z)}(\coprod_{j}W_{j})$ exists and there is a
natural isomorphism
\begin{equation}
\biggl(\coprod_{i}U_{i}\biggr)\boxtimes_{Q(z)}
\biggl(\coprod_{j}W_{j}\biggr)\stackrel{\sim}
{\rightarrow} \coprod_{i,j}U_{i}\boxtimes_{Q(z)}W_{j}.\hspace{2em} \Box
\end{equation}
\end{propo}

Now consider $V$-modules $W_{1}$, $W_{2}$ and $W_{3}$ and suppose that
$\dim {\cal M}_{W_{1}W_{2}}^{W_{3}}<\infty$. The natural
evaluation map
\begin{eqnarray}
W_{1}\otimes W_{2}\otimes {\cal M}^{W_{3}}_{W_{1}W_{2}}&\to& \overline{W}_{3}
\nno\\
w_{(1)}\otimes w_{(2)}\otimes F&\mapsto& F(w_{(1)}\otimes w_{(2)})
\end{eqnarray}
gives a natural map
\begin{equation}
{\cal F}^{W_{3}}_{W_{1}W_{2}}: W_{1}\otimes W_{2}\to
\mbox{\rm Hom}({\cal M}_{W_{1}W_{2}}^{W_{3}}, \overline{W}_{3})=
({\cal M}^{W_{3}}_{W_{1}W_{2}})^{*}\otimes \overline{W}_{3}.
\end{equation}
Also,
$({\cal M}^{W_{3}}_{W_{1}W_{2}})^{*}\otimes W_{3}$ is a $V$-module
(with finite-dimensional weight spaces) in the obvious way, and the map
${\cal F}^{W_{3}}_{W_{1}W_{2}}$ is clearly a $Q(z)$-intertwining map, where we
make the identification
\begin{equation}
({\cal M}^{W_{3}}_{W_{1}W_{2}})^{*}\otimes \overline{W}_{3}
=\overline{({\cal M}^{W_{3}}_{W_{1}W_{2}})^{*}\otimes W_{3}}.
\end{equation}
This gives us a natural $Q(z)$-product. Moreover, we have a natural linear
injection
\begin{eqnarray}
i: {\cal M}^{W_{3}}_{W_{1}W_{2}}&\to &
\mbox{\rm Hom}_{V}(({\cal M}^{W_{3}}_{W_{1}W_{2}})^{*}\otimes W_{3},
W_{3})\nno\\
F&\mapsto &(f\otimes w_{(3)}\mapsto f(F)w_{(3)})
\end{eqnarray}
which is an isomorphism if $W_{3}$ is irreducible, since in this
case,
$$\mbox{\rm Hom}_{V}(W_{3}, W_{3})\simeq {\Bbb C}$$
 (see \cite{FHL}, Remark 4.7.1).
On the other hand, the natural map
\begin{eqnarray}
h:\mbox{\rm Hom}_{V}(({\cal M}^{W_{3}}_{W_{1}W_{2}})^{*}
\otimes W_{3}, W_{3})&\to &
{\cal M}^{W_{3}}_{W_{1}W_{2}}\nno\\
\eta&\mapsto &\overline{\eta}\circ {\cal F}^{W_{3}}_{W_{1}W_{2}}
\end{eqnarray}
given by composition clearly satisfies the condition that
\begin{equation}
h(i(F))=F,
\end{equation}
so that if $W_{3}$ is irreducible, the maps $h$ and $i$ are mutually inverse
isomorphisms and we have the universal property that for any $F\in
{\cal M}^{W_{3}}_{W_{1}W_{2}}$, there exists a unique $\eta$ such that
\begin{equation}
F=\overline{\eta}\circ {\cal F}^{W_{3}}_{W_{1}W_{2}}
\end{equation}
(cf. Definition 4.4).

 Now we consider a special but important class of
vertex operator algebras satisfying certain finiteness and
semisimplicity conditions.
\begin{defi}
{\rm A  vertex operator algebra $V$ is {\it rational} if it
satisfies the following conditions:
\begin{enumerate}
\item There are only finitely many irreducible $V$-modules (up to equivalence).
\item Every $V$-module is completely reducible (and is in particular a
{\it finite} direct sum of irreducible modules).
\item All the fusion rules for $V$ are finite (for triples of irreducible
modules and hence arbitrary modules).
\end{enumerate}
}
\end{defi}

The next result shows that tensor products exist for the category of modules
for a rational vertex operator algebra. Note that there is no need to assume
that $W_{1}$ and $W_{2}$ are irreducible in the formulation or proof, but
by Proposition 4.11, the case in which $W_{1}$ and $W_{2}$ are
irreducible is in fact sufficient, and the
tensor product operation is canonically described using only the spaces
of intertwining maps among triples of {\it irreducible} modules.

\begin{propo}
Let $V$ be rational and let $W_{1}$, $W_{2}$ be $V$-modules. Then
$(W_{1}\boxtimes_{Q(z)}W_{2}, Y_{Q(z)}; \boxtimes_{Q(z)})$ exists, and in fact
\begin{equation}
W_{1}\boxtimes_{Q(z)}W_{2}=\coprod_{i=1}^{k}
({\cal M}^{M_{i}}_{W_{1}W_{2}})^{*}\otimes M_{i},
\end{equation}
where $\{ M_{1}, \dots, M_{k}\}$ is a set of representatives of the
equivalence classes of irreducible $V$-modules,  and the right-hand side of
(4.41) is equipped with the $V$-module and $Q(z)$-product structure indicated
above. That is,
\begin{equation}
\boxtimes_{Q(z)}=\sum_{i=1}^{k}{\cal F}^{M_{i}}_{W_{1}W_{2}}.
\end{equation}
\end{propo}
\pf
{From} the comments above and the definitions, it is clear that we have a
$Q(z)$-product. Let $(W_{3}, Y_{3}; F)$ be any $Q(z)$-product. Then
$W_{3}=\coprod_{j}U_{j}$ where $j$ ranges through a finite set and each
$U_{j}$ is irreducible. Let $\pi_{j}: W_{3}\to U_{j}$ denote the $j$-th
projection. A module map $\eta:\coprod_{i=1}^{k}
({\cal M}^{M_{i}}_{W_{1}W_{2}})^{*}\otimes M_{i}\to W_{3}$ amounts to
module maps
$$\eta_{ij}:  ({\cal M}^{M_{i}}_{W_{1}W_{2}})^{*}\otimes M_{i}\to U_{j}$$
for each $i$ and $j$ such that $U_{j}\simeq M_{i}$, and
$F=\overline{\eta}\circ \boxtimes_{Q(z)}$ if and only if
$$\overline{\pi}_{j}\circ F=\overline{\eta}_{ij}\circ
{\cal F}^{M_{i}}_{W_{1}W_{2}}$$
for each $i$ and $j$, the bars having the obvious meaning. But
$\overline{\pi}_{j}\circ F$ is a $Q(z)$-intertwining map of type
${U_{j}}\choose {W_{1}W_{2}}$, and so
$\overline{\iota}\circ \overline{\pi}_{j}\circ F\in
{\cal M}^{M_{i}}_{W_{1}W_{2}}$,
where $\iota: U_{j}\stackrel{\sim}{\to}M_{i}$ is a fixed isomorphism.
Denote this map by $\tau$. Thus what we finally want is a unique module map
$$\theta: ({\cal M}^{M_{i}}_{W_{1}W_{2}})^{*}\otimes M_{i}\to M_{i}$$
such that
$$\tau=\overline{\theta}\circ {\cal F}^{M_{i}}_{W_{1}W_{2}}.$$
But we in fact have such a unique $\theta$, by (4.39)--(4.40).\epf

\begin{rema}
By combining Proposition 4.13 with Proposition 4.7 or Proposition 4.9, we can
express $W_{1}\boxtimes_{Q(z)} W_{2}$ in terms of
${\cal V}^{W'_{1}}_{M'_{i}W_{2}}$ or ${\cal V}^{M_{i}}_{W_{1}W_{2}}$ in
place of ${\cal M}^{M_{i}}_{W_{1}W_{2}}$.
\end{rema}

The construction in Proposition 4.13 is tautological, and we view the argument
as essentially an existence proof. In the next two sections, we present
``first and second constructions'' of a $Q(z)$-tensor product.

\renewcommand{\theequation}{\thesection.\arabic{equation}}
\renewcommand{\therema}{\thesection.\arabic{rema}}
\setcounter{equation}{0}
\setcounter{rema}{0}

\section{First construction of  $Q(z)$-tensor product}

Here and in the next section, we give two constructions of a
$Q(z)$-tensor product of two modules for a vertex operator algebra $V$, in the
presence of a certain hypothesis which holds in case $V$ is rational.
In this section, we first define an action of
$V \otimes \iota_{+}{\Bbb C}[t,t^{- 1},(z+t)^{-1}]$ on $(W_1 \otimes
W_2)^*$ motivated by the definition (4.4) of $Q(z)$-intertwining
map. We establish some basic properties of this action, deferring the
proof of a commutator formula (Proposition 5.2) to Part II.
Then we take the sum of all ``compatible modules'' in $(W_1
\otimes W_2)^*$. Under the assumption that this sum is again
a module, we construct the
$Q(z)$-tensor product as its contragredient module equipped with
the restriction to $W_{1}\otimes W_{2}$ of the adjoint of the
embedding map of this sum in $(W_1 \otimes W_2)^*$. In the next
section we observe that every element in the sum of compatible
modules in $(W_1 \otimes W_2)^*$ satisfies a certain set of
conditions, and we
show that, modulo two important results stated there but whose proofs are
deferred to Part II, the
subspace of $(W_1 \otimes W_2)^*$ consisting of all the elements satisfying
these conditions is equal to this sum of compatible modules.
In this way we obtain another
construction of the $Q(z)$-tensor product.

Fix a nonzero complex number $z$ and $V$-modules $(W_{1}, Y_{1})$ and
$(W_{2}, Y_{2})$ as before.  We first want to define a linear action
of $V \otimes \iota_{+}{\Bbb C}[t,t^{- 1},(z+t)^{-1}]$ on $(W_1
\otimes W_2)^*$, that is, a linear map $$\tau_{Q(z)}: V\otimes
\iota_{+}{\Bbb C}[t, t^{-1}, (z+t)^{-1}]\to
\mbox{\rm End}\;(W_{1}\otimes W_{2})^{*}.$$
Recall the maps
$$\tau_{W_{i}}:  V\otimes {\Bbb C}((t))\to \mbox{\rm End}\;
W_{i},\;\;\;i=1, 2,$$
{from} (3.2). We
define $\tau_{Q(z)}$ by
\begin{equation}
(\tau_{Q(z)}(\xi)\lambda)(w_{(1)}\otimes w_{(2)})
=\lambda(\tau_{W_{1}}(T_{-z}^{*}\xi)w_{(1)}\otimes w_{(2)})
-\lambda(w_{(1)}\otimes \tau_{W_{2}}(T_{-z}^{+}\xi)w_{(2)})
\end{equation}
for $\xi\in V\otimes \iota_{+}{\Bbb C}[t, t^{-1}, (z+t)^{-1}]$,
$\lambda\in (W_{1}\otimes W_{2})^{*}$, $w_{(1)}\in W_{1}$, $w_{(2)}\in W_{2}$.
Using (3.12)--(3.13), (3.60) and Lemma 3.1, we see that
the definition (5.1) can be written using
generating functions as:
\bea
\lefteqn{\left(\tau_{Q(z)}
\left(z^{-1}\delta\left(\frac{x_{1}-x_{0}}{z}\right)
Y_{t}(v, x_{0})\right)\lambda\right)(w_{(1)}\otimes w_{(2)})}\nonumber\\
&&=x^{-1}_{0}\delta\left(\frac{x_{1}-z}{x_{0}}\right)
\lambda(Y_{1}^{*}(v, x_{1})w_{(1)}\otimes w_{(2)})\nonumber\\
&&\hspace{2em}-x_{0}^{-1}\delta\left(\frac{z-x_{1}}{-x_{0}}\right)
\lambda(w_{(1)}\otimes Y_{2}(v, x_{1})w_{(2)}).
\eea

Write
\be
Y'_{Q(z)}(v, x)=\tau_{Q(z)}(Y_{t}(v, x)).
\ee
Using (2.6) and the fundamental property of the formal $\delta$-function,
we have
\begin{eqnarray}
\lefteqn{(Y'_{Q(z)}(v,x_0)\lambda)(w_{(1)} \otimes
w_{(2)})=}\nno\\
&&= \res_{x_1} x^{-1}_0 \delta\left(\frac{x_1-z}{x_0}\right)
\lambda(Y^*_1(v,x_1)w_{(1)} \otimes w_{(2)})\nno\\
&&\hspace{2em}-
\res_{x_1}x^{-1}_0 \delta \left(\frac{z-x_1}{-x_0}\right)\lambda(w_{(1)}
\otimes Y_2(v,x_1)w_{(2)})\nno\\
&&= \lambda(Y^*_1(v,x_0 + z)w_{(1)}
\otimes w_{(2)}) \nno\\
&&\hspace{2em}-
\res_{x_1}x^{-1}_0 \delta \left(\frac{z-x_1}{-x_0}\right)\lambda(w_{(1)}
\otimes Y_2(v,x_1)w_{(2)}),
\end{eqnarray}
where we have used the notation $\res_{x_{1}}$, which means taking the
coefficient of $x_{1}$ in a formal series.
We have the following results for $Y'_{Q(z)}$:

\begin{propo}
The action $Y'_{Q(z)}$ has the property
\begin{equation}
Y'_{Q(z)}({\bf 1}, x)=1,
\end{equation}
where $1$ on the right-hand side is the identity map of
$(W_{1}\otimes W_{2})^{*}$, and the $L(-1)$-derivative property
\begin{equation}
\frac{d}{dx}Y'_{Q(z)}(v, x)=Y'_{Q(z)}(L(-1)v, x)
\end{equation}
for $v\in V$.
\end{propo}
\pf
{From} (5.4), (3.20) and (2.7),
\begin{eqnarray}
\lefteqn{(Y({\bf 1},x)\lambda)(w_{(1)} \otimes
w_{(2)}) =}\nno\\
&& =\res_{x_1}x^{-1} \delta\left(\frac{x_1-
z}{x}\right)\lambda(w_{(1)} \otimes w_{(2)})\nno\\
&& \hspace{2em}-\res_{x_1}x^{-1}
\delta\left(\frac{z-x_1}{-x}\right)\lambda(w_{(1)} \otimes w_{(2)})\nno\\
&&=
\res_{x_1}x^{-1}_1 \delta\left(\frac{z+x}{x_1}\right)\lambda(w_{(1)}
\otimes w_{(2)})\nno\\
&&= \lambda(w_{(1)} \otimes
w_{(2)}),
\end{eqnarray}
proving (5.5). We now prove the $L(-1)$-derivative property.
{From} (5.4),
\begin{eqnarray}
\lefteqn{\left(\left(\frac{d}{dx}
Y'_{Q(z)}(v,x)\right)\lambda\right)(w_{(1)} \otimes w_{(2)})=}\nno\\
&&= \frac{d}{dx}
\lambda(Y^*_1(v,x + z)w_{(1)} \otimes w_{(2)})\nno\\
&&\hspace{2em}
-\res_{x_1}\left(\frac{d}{dx} z^{-1}\delta\left(\frac{-x + x_1}{z}\right)
\right)\lambda(w_{(1)}
\otimes Y_2(v,x_1)w_{(2)}).
\end{eqnarray}
Note that
for any formal Laurent series $f(x)$, we have
\begin{equation}
\frac{d}{dx}f\left(\frac{-x+x_{1}}{z}\right)
=-\frac{d}{dx_{1}}f\left(\frac{-x+x_{1}}{z}\right)
\end{equation}
and if $f(x)$ involves only finitely many negative powers of $x$,
\begin{equation}
\res_{x_{1}}\left(\frac{d}{dx_{1}}z^{-1}\delta\left(
\frac{-x+x_{1}}{z}\right)\right)f(x_{1})=
-\res_{x_{1}}z^{-1}\delta\left(
\frac{-x+x_{1}}{z}\right)\frac{d}{dx_{1}}f(x_{1})
\end{equation}
(since the residue of a derivative is $0$).
{From} (3.22) and the $L(-1)$-derivative property for the contragredient module
$W'_{1}$, we have
$$\frac{d}{dx} Y^*_{1}(v,x) = Y^*_{1}(L(-1)v,x).$$
Thus the right-hand side of (5.8) is equal to
\begin{eqnarray}
\lefteqn{\lambda(Y^*_1(L(-
1)v,x+z)w_{(1)} \otimes w_{(2)})}\nno\\
&&\hspace{2em}-
\res_{x_1}z^{-1}\delta\left(
\frac{-x+x_{1}}{z}\right)\frac{d}{dx_1}\lambda(w_{(1)} \otimes
Y_2(v,x_1)w_{(2)})\nno\\
&&=  \lambda(Y^*_1(L(-1)v,x+z)w_{(1)}
\otimes w_{(2)})\nno\\
&&\hspace{2em}-  \res_{x_1}z^{-1}\delta\left(
\frac{-x+x_{1}}{z}\right)\lambda(w_{(1)}
\otimes Y_2(L(-1)v,x_1)w_{(2)})\nno\\
&&=  (Y'_{Q(z)}(L(-1)v,x)\lambda)(w_{(1)}
\otimes w_{(2)}),
\end{eqnarray}
completing the proof. \epf

\begin{propo}
The action $Y'_{Q(z)}$ satisfies the commutator formula for vertex operators,
that is, on
$(W_{1}\otimes W_{2})^{*}$,
\bea
\lefteqn{[Y'_{Q(z)}(v_{1}, x_{1}), Y'_{Q(z)}(v_{2}, x_{2})]=}\nno\\
&&=\res_{x_{0}}x_{2}^{-1}\delta\left(\frac{x_{1}-x_{0}}{x_{2}}\right)
Y'_{Q(z)}(Y(v_{1}, x_{0})v_{2}, x_{2})
\eea
for $v_{1}, v_{2}\in V$.
\end{propo}

The proof of this proposition will be given in Part II.

{From} these results and the relation (2.22), we see that the
coefficient operators of $Y'_{Q(z)}(\omega, x)$ satisfy the Virasoro
algebra commutator relations, that is, writing
\begin{equation}
Y'_{Q(z)}(\omega, x)=\sum_{n\in {\Bbb Z}}L'_{Q(z)}(n)x^{-n-2},
\end{equation}
we have
\begin{equation}
[L'_{Q(z)}(m), L'_{Q(z)}(n)]
=(m-n)L'_{Q(z)}(m+n)+{\displaystyle\frac{1}{12}}
(m^{3}-m)\delta_{m+n,0}c.
\end{equation}
We call the eigenspaces of the operator $L'_{Q(z)}(0)$ the {\it weight
subspaces} or {\it
homogeneous subspaces} of $(W_{1}\otimes W_{2})^{*}$, and we have the
corresponding notions of {\it weight vector} (or {\it homogeneous
vector}) and  {\it weight}. When there is no confusion,
we shall simply write $L'_{Q(z)}(n)$ as $L(n)$.

Let
$W_{3}$ be another $V$-module.
Note that $V\otimes \iota_{+}{\Bbb C}[t, t^{-1}, (z+t)^{-1}]$ acts on
$W'_{3}$ in the obvious way. The following  result, which follows
immediately {from} the definitions (4.4) and (5.2),
provides further motivation for the definition of our action on
$(W_{1}\otimes W_{2})^{*}$:

\begin{propo}
Under the natural isomorphism
\begin{equation}
\mbox{\rm Hom}(W'_{3}, (W_{1}\otimes
W_{2})^{*})\stackrel{\sim}{\to}\mbox{\rm Hom}(W_{1}\otimes W_{2},
\overline{W}_{3}),
\end{equation}
the maps in $\mbox{\rm Hom}(W'_{3}, (W_{1}\otimes
W_{2})^{*})$ intertwining the two actions of
$V \otimes \iota_{+}{\Bbb C}[t,t^{-
1},(z+t)^{-1}]$ on $W'_{3}$ and
$(W_{1}\otimes W_{2})^{*}$ correspond exactly to the
$Q(z)$-intertwining maps of type ${W_{3}}\choose {W_{1}W_{2}}$. \epf
\end{propo}

\begin{rema}
{\rm Combining the last result with Proposition 4.7, we see that
the maps in $\mbox{\rm Hom}(W'_{3}, (W_{1}\otimes
W_{2})^{*})$ intertwining the two actions on $W'_{3}$ and
$(W_{1}\otimes W_{2})^{*}$ also correspond exactly to the
intertwining operators of type
${W_{1}'}\choose {W'_{3}\;W_{2}}$. In particular, given any integer $p$,
the map
$F'_{{\cal Y}, p}: W'_{3}\to (W_{1}\otimes W_{2})^{*}$ defined by
\begin{equation}
F'_{{\cal Y}, p}(w'_{(3)})(w_{(1)}\otimes w_{(2)})=\bra w_{(1)},
{\cal Y}(w'_{(3)}, e^{l_{p}(z)})w_{(2)}\ket_{W'_1}
\end{equation}
(recall (4.9))  intertwines the
actions  of $V\otimes \iota_{+}{\Bbb C}[t, t^{-1}, (z+t)^{-1}]$ on
$W'_{3}$ and
$(W_{1}\otimes W_{2})^{*}$.}
\end{rema}

Suppose that $G\in \mbox{\rm Hom}(W'_{3}, (W_{1}\otimes
W_{2})^{*})$ intertwines the two actions as in Proposition 5.3.
Then for $w'_{(3)}\in W'_{3}$,
\begin{eqnarray}
\lefteqn{\tau_{Q(z)}\left(z^{-1}\delta\left(\frac{x_{1}-x_{0}}{z}\right)
Y_{t}(v, x_{0})\right)G(w'_{(3)})=}\nno\\
&&=G\left(\tau_{W'_{3}}\left(z^{-1}\delta\left(\frac{x_{1}-x_{0}}{z}\right)
Y_{t}(v, x_{0})\right)w'_{(3)}\right)\nno\\
&&=G\left(z^{-1}\delta\left(\frac{x_{1}-x_{0}}{z}\right)
Y'_{3}(v, x_{0})w'_{(3)}\right)\nno\\
&&=z^{-1}\delta\left(\frac{x_{1}-x_{0}}{z}\right)G(Y'_{3}(v,
x_{0})w'_{(3)})\nno\\
&&=z^{-1}\delta\left(\frac{x_{1}-x_{0}}{z}\right)
Y'_{Q(z)}(v, x_{0})G(w'_{(3)}).
\end{eqnarray}
Thus $G(w'_{(3)})$ satisfies the following nontrivial and subtle condition
on $\lambda
\in (W_{1}\otimes W_{2})^{*}$: The formal Laurent series $Y'_{Q(z)}(v,
x_{0})\lambda$ involves only finitely many negative powers of $x_{0}$
and
\begin{eqnarray}
\lefteqn{\tau_{Q(z)}\left(z^{-1}\delta\left(\frac{x_{1}-x_{0}}{z}\right)
Y_{t}(v, x_{0})\right)\lambda=}\nno\\
&&=z^{-1}\delta\left(\frac{x_{1}-x_{0}}{z}\right)
Y'_{Q(z)}(v, x_{0})\lambda  \;\;\;\;\; \mbox{\rm for all}\;\;v\in V.
\end{eqnarray}
(Note that the two sides are not {\it a priori} equal for general
$\lambda\in (W_{1}\otimes W_{2})^{*}$.)
We call this the {\it compatibility  condition} on
$\lambda\in (W_{1}\otimes W_{2})^{*}$, for the action
$\tau_{Q(z)}$.

Let $W$ be a subspace of $(W_{1}\otimes W_{2})^{*}$.  We say that $W$
is {\it compatible for $\tau_{Q(z)}$} if every element of $W$
satisfies the compatibility condition. Also, we say that $W$ is (${\Bbb
C}$-){\it graded} if it is ${\Bbb C}$-graded by its weight subspaces,
and that $W$ is a $V$-{\it module} (respectively, {\it
generalized module}) if $W$ is graded and is a module (respectively,
generalized module) when equipped with this grading and with the action of
$Y'_{Q(z)}(\cdot, x)$ (recall Definition 2.11). A sum of compatible
modules or generalized modules is clearly a generalized module. The
weight subspace of a subspace $W$ with weight $n\in {\Bbb C}$ will be
denoted $W_{(n)}$.

Given $G$ as above, it is clear that $G(W'_{3})$ is a $V$-module since $G$
intertwines the two actions of $V\otimes {\Bbb C}[t, t^{-1}]$. We have in
fact established that $G(W'_{3})$ is in addition a compatible $V$-module since
$G$ intertwines the full actions. Moreover, if
$H\in \mbox{\rm Hom}(W'_{3}, (W_{1}\otimes W_{2})^{*})$ intertwines the two
actions of $V\otimes {\Bbb C}[t, t^{-1}]$, then $H$ intertwines the two actions
of $V\otimes \iota_{+}{\Bbb C}[t, t^{-1}, (z+t)^{-1}]$ if and only if the
$V$-module $H(W'_{3})$ is compatible.

Define
\be
W_{1}\hboxtr_{Q(z)}W_{2}=\sum_{W\in {\cal W}_{Q(z)}}W =\bigcup_{W\in
{\cal W}_{Q(z)}} W\subset
(W_{1}\otimes W_{2})^{*},
\ee
where ${\cal W}_{Q(z)}$ is the set all compatible
modules for $\tau_{Q(z)}$ in $(W_{1}\otimes W_{2})^{*}$.
Then $W_{1}\hboxtr_{Q(z)}W_{2}$
is a compatible generalized module and coincides with the sum (or
union) of the
images $G(W'_{3})$ of modules $W'_{3}$ under the maps $G$ as above.
Moreover, for any $V$-module $W_{3}$ and any map
$H: W'_{3}\to W_{1}\hboxtr_{Q(z)} W_{2}$ of generalized modules,
$H(W'_{3})$ is compatible and hence $H$ intertwines the two actions
of $V\otimes \iota_{+}{\Bbb C}[t, t^{-1}, (z+t)^{-1}]$. Thus we have:
\begin{propo}
The subspace $W_{1}\hboxtr_{Q(z)}W_{2}$ of $(W_{1}\otimes W_{2})^{*}$ is
a generalized module with the following
property: Given  any $V$-module $W_{3}$,
there is a natural linear isomorphism determined by (5.15) between the space of
all $Q(z)$-intertwining maps of type ${W_{3}}\choose {W_{1}\;W_{2}}$
and the space of all maps of generalized modules {from} $W'_{3}$ to
$W_{1}\hboxtr _{Q(z)}W_{2}$.  \epf
\end{propo}

\begin{propo}
Let $V$ be a rational vertex operator algebra and $W_{1}$, $W_{2}$ two
$V$-modules. Then $W_{1}\hboxtr _{Q(z)}W_{2}$ is a module.
\end{propo}
\pf
 Since $W_{1}\hboxtr _{Q(z)}W_{2}$ is the sum of all compatible
modules for $\tau_{Q(z)}$ in $(W_{1}\otimes W_{2})^{*}$ and since by
assumption every module is completely reducible, the generalized $V$-module
$W_{1}\hboxtr_{Q(z)}W_{2}$ is a direct sum of irreducible modules.
If it is an infinite direct sum, it must include infinitely many
copies of at least one irreducible $V$-module, say, $W_{3}$,
since a rational vertex operator algebra
has only finitely many irreducible modules.  {From} Proposition 5.5, the
space  of $Q(z)$-intertwining maps of
type ${W'_{3}}\choose {W_{1}\;W_{2}}$ must be infinite-dimensional, and by
Proposition 4.7,
this contradicts the assumed finiteness of the fusion rules. Thus
$W_{1}\hboxtr_{Q(z)} W_{2}$ is a finite direct sum of irreducible
modules and hence is a module.
\epfv

Now we assume that $W_{1}\hboxtr_{Q(z)} W_{2}$ is a module (which
occurs if $V$ is rational, by
the last proposition).  In this case, we define a $V$-module
$W_{1}\boxtimes_{Q(z)} W_{2}$ by
\begin{equation}
W_{1}\boxtimes_{Q(z)} W_{2}=(W_{1}\hboxtr_{Q(z)}W_{2})'
\end{equation}
($\hboxtr'=\boxtimes !$) and we write the corresponding
action as $Y_{Q(z)}$. Applying Proposition 5.5 to the special module
$W_{3}=W_{1}\boxtimes_{Q(z)} W_{2}$ and the identity map $W'_{3}\to
W_{1}\hboxtr_{Q(z)} W_{2}$ (recall Theorem 2.10), we obtain using
(5.15) a canonical $Q(z)$-intertwining map of type
${W_{1}\boxtimes_{Q(z)} W_{2}}\choose {W_{1}W_{2}}$, which we denote
\begin{eqnarray}
\boxtimes_{Q(z)}: W_{1}\otimes W_{2}&\to &
\overline{W_{1}\boxtimes_{Q(z)} W_{2}}\nno\\
w_{(1)}\otimes  w_{(2)}&\mapsto& w_{(1)} \boxtimes_{Q(z)}w_{(2)}.
\end{eqnarray}
This is the unique linear map such that
\begin{equation}
\langle \lambda, w_{(1)} \boxtimes_{Q(z)}
w_{(2)}\rangle_{W_{1}\boxtimes_{Q(z)} W_{2}}
=\lambda(w_{(1)}\otimes w_{(2)})
\end{equation}
for all $w_{(1)}\in W_{1}$, $w_{(2)}\in W_{2}$ and
$\lambda\in W_{1}\hboxtr_{Q(z)} W_{2}$. Moreover, we have:

\begin{propo}
The $Q(z)$-product $(W_{1}\boxtimes_{Q(z)} W_{2}, Y_{Q(z)};
\boxtimes_{Q(z)})$
is a $Q(z)$-ten- sor product of $W_{1}$ and $W_{2}$.
\end{propo}
{\it Proof}\hspace{2ex} Let
$(W_{3}, Y_{3}; F)$ be a $Q(z)$-product of $W_{1}$ and $W_{2}$.
By Proposition 5.5, there is a unique $V$-module map
$$\eta': W'_{3}\to W_{1}\hboxtr_{Q(z)} W_{2}$$
such that
$$
\langle w'_{(3)}, F(w_{(1)}\otimes w_{(2)})\rangle_{W_3}
=\eta'(w'_{(3)})(w_{(1)}\otimes w_{(2)}) $$ for any $w_{(1)}\in
W_{1}$, $w_{(2)}\in W_{2}$ and $w'_{(3)}\in W'_{3}$. But by (5.22),
this equals $$\langle \eta'(w'_{(3)}), w_{(1)}\boxtimes_{Q(z)}
w_{(2)}\rangle_{W_{1}\boxtimes_{Q(z)} W_{2}} =\langle w'_{(3)},
\overline{\eta}(w_{(1)}\boxtimes_{Q(z)} w_{(2)})\rangle_{W_3},$$ where
$\eta\in \mbox{\rm Hom}_{V}(W_{1}\boxtimes_{Q(z)} W_{2}, W_{3})$ and
$\eta'$ are mutually adjoint maps. In particular, there is a unique
$\eta$ such that $$\langle w'_{(3)}, F(w_{(1)}\otimes
w_{(2)})\rangle_{W_3} =\langle w'_{(3)},
\overline{\eta}(w_{(1)}\boxtimes_{Q(z)} w_{(2)})\rangle_{W_3},$$ i.e.,
such that $$F=\overline{\eta}\circ \boxtimes_{Q(z)}: W_{1}\otimes
W_{2}\to
\overline{W}_{3},$$
and this establishes the desired universal property.\epfv

More generally, dropping the assumption that $W_{1}\hboxtr_{Q(z)}
W_{2}$ is a module, we have:

\begin{propo}
The $Q(z)$-tensor product of $W_{1}$ and $W_{2}$ exists (and is given
by (5.20)) if and only if $W_{1}\hboxtr_{Q(z)} W_{2}$ is a module.
\end{propo}
\pf
It is sufficient to show that if the $Q(z)$-tensor product exists, then
$W_{1}\hboxtr_{Q(z)} W_{2}$ is a module.
Consider the module
$$W_{0}=(W_{1}\boxtimes_{Q(z)} W_{2})'.$$
Applying Proposition 5.5 to the $Q(z)$-product $W_{1}\boxtimes_{Q(z)}
W_{2}$, we have a unique map
$$i: W_{0}\to W_{1}\hboxtr_{Q(z)} W_{2}$$
of generalized modules such that
$$i(w_{(0)})(w_{(1)}\otimes w_{(2)})=\langle w_{(0)},
w_{(1)}\boxtimes_{Q(z)} w_{(2)}\rangle_{W_{1}\boxtimes_{Q(z)} W_{2}}$$
for $w_{(0)}\in W_{0}$, $w_{(1)}\in W_{1}$ and $w_{(2)}\in W_{2}$. It
suffices to show that $i$ is a surjection.

Let $W\in {\cal W}_{Q(z)}$ (recall (5.19)) and set $W_{3}=W'.$ By
Proposition 5.5, the injection $W'_{3}\hookrightarrow
W_{1}\hboxtr_{Q(z)} W_{2}$ induces a unique $Q(z)$-intertwining map
$F$ of type ${W'}\choose {W_{1}W_{2}}$ such that
$$w(w_{(1)}\otimes w_{(2)})=\langle w, F(w_{(1)}\otimes
w_{(2)})\rangle_{W'}$$
for $w\in W$, $w_{(1)}\in W_{1}$ and $w_{(2)}\in W_{2}$. But by the
universal property of $W_{1}\boxtimes_{Q(z)} W_{2}$, there is a unique
module map $\eta': W_{1}\boxtimes_{Q(z)} W_{2} \to W'$ such that
$F=\overline{\eta'} \circ \boxtimes_{Q(z)}$,
and hence a unique module map
$\eta: W\to W_{0}$ such that $$\langle \eta(w), w_{(1)}\boxtimes_{Q(z)}
w_{(2)}\rangle_{W_{1}\boxtimes_{Q(z)} W_{2}}=\langle w, F(w_{(1)}\otimes
w_{(2)})\rangle_{W'}.$$ Thus
\begin{eqnarray*}
w(w_{(1)}\otimes w_{(2)})&=&\langle \eta(w), w_{(1)}\boxtimes_{Q(z)}
w_{(2)}\rangle_{W_{1}\boxtimes_{Q(z)} W_{2}}\\
&=&i(\eta(w))(w_{(1)}\otimes w_{(2)})
\end{eqnarray*}
and so $w=i(\eta(w))$ for all $w\in W$, showing that $W$ lies in the
image of the map $i$ and hence that $i$ is surjective.\epf

\renewcommand{\theequation}{\thesection.\arabic{equation}}
\renewcommand{\therema}{\thesection.\arabic{rema}}
\setcounter{equation}{0}
\setcounter{rema}{0}

 \section{Second construction of  $Q(z)$-tensor product}

Let $V$ be a vertex operator algebra and $W_{1}$, $W_{2}$ two $V$-modules.
{From} the definition (5.19) of $W_{1}\hboxtr_{Q(z)} W_{2}$,
 we see that any element of
$W_{1}\hboxtr_{Q(z)} W_{2}$ is an element $\lambda$
of $(W_{1}\otimes W_{2})^{*}$ satisfying:

\begin{description}
\item[The compatibility condition] (recall (5.18)){\bf :}
{\bf (a)} The  {\it lower
truncation condition}:
For all $v\in V$, the formal Laurent series $Y'_{Q(z)}(v, x)\lambda$
involves only finitely many negative
powers of $x$.

{\bf (b)} The following formula holds:
\begin{eqnarray}
\lefteqn{\tau_{Q(z)}\left(z^{-1}\delta\left(\frac{x_{1}-x_{0}}{z}\right)
Y_{t}(v, x_{0})\right)\lambda=}\nno\\
&&=z^{-1}\delta\left(\frac{x_{1}-x_{0}}{z}\right)
Y'_{Q(z)}(v, x_{0})\lambda  \;\;\;\;\; \mbox{\rm for all}\;\;v\in V.
\end{eqnarray}

\item[The local grading-restriction  condition:]
{\bf (a)} The {\it grading condition}:
$\lambda$ is a (finite) sum of
weight vectors of $(W_{1}\otimes W_{2})^{*}$.

{\bf (b)} Let $W_{\lambda}$ be the smallest subspace of $(W_{1}\otimes
W_{2})^{*}$ containing $\lambda$ and stable under the component
operators $\tau_{Q(z)}(v\otimes t^{n})$ of the operators $Y'_{Q(z)}(v,
x)$ for $v\in V$, $n\in {\Bbb Z}$. Then the weight spaces
$(W_{\lambda})_{(n)}$, $n\in {\Bbb C}$, of the (graded) space
$W_{\lambda}$ have the properties
\begin{eqnarray}
&\mbox{\rm dim}\ (W_{\lambda})_{(n)}<\infty \;\;\;\mbox{\rm for}\
n\in {\Bbb C},&\\
&(W_{\lambda})_{(n)}=0 \;\;\;\mbox{\rm for $n$ whose real part is
sufficiently small.}&
\end{eqnarray}
\end{description}

Note that the set of the elements of $(W_{1}\otimes W_{2})^{*}$
satisfying any one of the lower truncation condition, the
compatibility condition, the grading condition or the local
grading-restriction condition forms a subspace.

In Part II, we shall prove the following
two basic results:

\begin{theo}
Let $\lambda$ be an element of $(W_{1}\otimes W_{2})^{*}$ satisfying
the compatibility condition. Then when acting on $\lambda$, the
Jacobi identity for $Y'_{Q(z)}$ holds, that is,
\begin{eqnarray}
\lefteqn{x_{0}^{-1}\delta
\left({\displaystyle\frac{x_{1}-x_{2}}{x_{0}}}\right)Y'_{Q(z)}(u, x_{1})
Y'_{Q(z)}(v, x_{2})\lambda}\nno\\
&&\hspace{2ex}-x_{0}^{-1} \delta
\left({\displaystyle\frac{x_{2}-x_{1}}{-x_{0}}}\right)Y'_{Q(z)}(v, x_{2})
Y'_{Q(z)}(u, x_{1})\lambda\nonumber \\
&&=x_{2}^{-1} \delta
\left({\displaystyle\frac{x_{1}-x_{0}}{x_{2}}}\right)Y'_{Q(z)}(Y(u, x_{0})v,
x_{2})\lambda
\end{eqnarray}
for $u, v\in V$.
\end{theo}

\begin{propo}
The subspace consisting of the elements of $(W_{1}\otimes W_{2})^{*}$
satisfying the compatibility condition is stable under the operators
$\tau_{Q(z)}(v\otimes t^{n})$ for $v\in V$ and $n\in {\Bbb Z}$, and
similarly for the subspace consisting of the elements satisfying the
local grading-restriction condition.
\end{propo}

These results give us another construction of $W_{1}\hboxtr_{Q(z)} W_{2}$:

\begin{theo}
The subspace of $(W_{1}\otimes W_{2})^{*}$ consisting of the
elements satisfying the compatibility
condition and the local grading-restriction condition, equipped with
$Y'_{Q(z)}$, is a generalized module and is equal to
$W_{1}\hboxtr_{Q(z)} W_{2}$.
\end{theo}
{\it Proof}\hspace{2ex} Let $W_{0}$ be the space of vectors satisfying
the two conditions. We have already observed that $W_{1}\hboxtr_{Q(z)}
W_{2}\subset W_{0}$, and it suffices to show that $W_{0}$ is
a generalized module which is a union
of compatible modules. But $W_{0}$ is a compatible generalized module
by Theorem 6.1 and Proposition 6.2, together with Proposition 5.1 and
formula (5.14), and every element of $W_{0}$ generates a compatible
module contained in $W_{0}$, by the local grading-restriction
condition.
\epfv

The following result follows immediately {from} Proposition 5.8, the
theorem above and the definition of $W_{1}\boxtimes_{Q(z)} W_{2}$:

\begin{corol}
The $Q(z)$-tensor product of $W_{1}$ and $W_{2}$ exists if and only if
the subspace of $(W_{1}\otimes W_{2})^{*}$ consisting of the elements
satisfying the compatibility condition and the local
grading-restriction condition, equipped with $Y'_{Q(z)}$,
is a module. In this case, this module coincides with the module $W_1
\hboxtr_{Q(z)} W_2$, and the contragredient module of this
module, equipped with the $Q(z)$-intertwining map $\boxtimes_{Q(z)}$,
 is a $Q(z)$-tensor product of $W_{1}$ and
$W_{2}$,  equal to the structure $(W_{1}\boxtimes_{Q(z)}
W_{2}, Y_{Q(z)};
\boxtimes_{Q(z)})$ constructed in Section 5.
\end{corol}

{From} this result and Propositions 5.6 and 5.7, we have:

\begin{corol}
Let $V$ be a rational vertex operator algebra and $W_{1}$, $W_{2}$ two
$V$-modules. Then the $Q(z)$-tensor product
$(W_{1}\boxtimes_{Q(z)} W_{2}, Y_{Q(z)};
\boxtimes_{Q(z)})$ may be constructed as described in Corollary 6.4.
\end{corol}

 {\small \sc Department of Mathematics, University of Pennsylvania,
Philadelphia, PA 19104}

 {\it Current address:} Department of Mathematics, Rutgers University,
New Brunswick, NJ 08903

{\em E-mail address}: yzhuang@math.rutgers.edu

\vskip 1em

{\small \sc Department of Mathematics, Rutgers University,
New Brunswick, NJ 08903}

{\em E-mail address}: lepowsky@math.rutgers.edu

\end{document}